\documentclass[useAMS,usenatbib]{mn2e}
\usepackage{graphicx,rotate,url,mathptmx,lscape,times}
\def\gtsim{\mathrel{\hbox{\rlap{\hbox{\lower4pt\hbox{$\sim$}}}\hbox{$>$}}}}
\def\lesssim{\mathrel{\hbox{\rlap{\hbox{\lower4pt\hbox{$\sim$}}}\hbox{$<$}}}}

\def\Msun{M$_{\odot}$}
\def\cm{{\rm\thinspace cm}}
\def\as{{\rm\thinspace arcsec}}
\def\erg{{\rm\thinspace erg}}

\def\K{{\rm\thinspace K}}

\def\km{{\rm\thinspace km}}
\def\kpc{{\rm\thinspace kpc}}

\def\Msun{\hbox{$\rm\thinspace M_{\odot}$}}

\def\s{{\rm\thinspace s}}
\def\ps{{\rm\thinspace s^{-1}}}

\def\yr{{\rm\thinspace yr}}
\def\ergpscmps{\hbox{$\erg\cm^{-2}\s^{-1}\,$}}
\def\ergpscmpspsas{\hbox{$\erg\cm^{-2}\s^{-1}\as^{-2}\,$}}

\def\kmps{\hbox{$\km\ps\,$}}

\def\pccm{\hbox{$\cm^{-3}\,$}}

\def\pccmK{\hbox{$\cm^{-3}\K$}}

\DeclareMathAlphabet{\vib}{OML}{cmm}{m}{it}

\title[Spitzer spectra of NGC 1275 and NGC 4696]{Discovery of 
  atomic and molecular mid infra-red emission lines in off-nuclear
  regions of NGC~1275 and NGC~4696 with
  the Spitzer Space Telescope}

\author[R.M. Johnstone, et al.]
       {\parbox[]{6.0in}
       {R.M. Johnstone$^1$\thanks{E-mail: rmj@ast.cam.ac.uk}, 
        N.A. Hatch$^{1,2}$, G.J. Ferland$^3$, A.C. Fabian$^1$,\\
        C.S. Crawford$^1$, and R.J. Wilman$^{4,5}$\\
        \footnotesize
        $^1$Institute of Astronomy, University of Cambridge, Madingley
        Road, Cambridge CB3 0HA\\
        $^2$Leiden Observatory, P.B. 9513, Leiden 2300 RA, The Netherlands\\
        $^3$Department of Physics, University of Kentucky, Lexington, KY 40506, USA\\
        $^4$Department of Physics, University of Durham, South Road, Durham, DH1 3LE\\
        $^5$Astrophysics, University of Oxford, Denys Wilkinson Building, Oxford, OX1 3RH}}

\date{
      Received }

\pagerange{\pageref{firstpage}--\pageref{lastpage}}
\pubyear{}

\begin{document}

\maketitle

\label{firstpage}

\begin{abstract}

\noindent
We present Spitzer high-resolution spectra of off-nuclear regions in
the central cluster galaxies NGC\,1275 and NGC\,4696 in the Perseus
and Centaurus clusters, respectively. Both objects are surrounded by
extensive optical emission-line filamentary nebulae, bright outer
parts of which are the targets of our observations.  The 10--37$\mu$m
spectra show strong pure rotational lines from molecular hydrogen
revealing a molecular component to the filaments which has an
excitation temperature of $\sim300-400$K.  The flux in the 0-0S(1)
molecular hydrogen line correlates well with the strength of the optical lines,
having about 3 per cent of the H$\alpha$+[NII] emission. The
11.3$\mu$m PAH feature is seen in some spectra.  Emission is also seen
from both low and high ionization fine structure lines.  Molecular
hydrogen cooler than $\sim$400K dominates the mass of the outer
filaments; the nebulae are predominantly molecular.

\end{abstract}


\section{Introduction}
\label{intro}

The massive central galaxy in many clusters is often surrounded by an
extensive emission-line nebulosity (e.g. \citealt{Cowie83,
Johnstone87, Heckman89, Crawford99}).  Such emission-line galaxies
are always at the centre of highly peaked X-ray emission from an
intracluster medium where the radiative cooling time of the
intracluster gas at the centre is less than 1~Gyr (\citealt{Fabian94,
Peres98, Bauer05}). Initially the nebulosity was studied in the
optical spectral region but strong UV and IR lines have since been
seen, together with dust which is inferred from the depletion of
calcium, Balmer line ratios and dust lanes which are present in some
cases. Recently, strong H$_2$ lines have been found (e.g.
\citealt{Jaffe97,Edge02} and refs therein) as well as emission lines
from the CO molecule (e.g. \citealt{Edge01, Salome03, Salome06}).

The ionization, excitation, origin and fate of
the filaments has been a long-standing puzzle. Most obvious sources of
ionization, such as an active nucleus, a radio source, shocks, young
stars and conduction from the surrounding hot gas have drawbacks in
one system or another.

XMM/RGS and Chandra X-ray spectra show that the intracluster gas in
many cluster cores has a range of temperatures decreasing to about a
factor of three below that of the outer hot gas, with very little
X-ray emitting gas cooler than this (\citealt{Peterson01}), indicating
that some form of heating (e.g. from the central radio source,
\citealt{Churazov02, Fabian03a}) is important and that it mostly
balances the radiative cooling. Heating is unlikely to completely
balance cooling over such an extended region, and so some residual
cooling flow is expected to occur. Optically emitting warm gas
filaments may be being dragged out from the central cluster galaxy by
buoyant bubbles (\citealt{Hatch06}) or may be a phase through which
some gas passes in cooling out from the intracluster medium. The wider
relevance of this issue is in the upper mass limit to the total
stellar component of massive galaxies, which may be controlled by
whatever heats cooling flows \citep{Fabian02b}. If radiative cooling
is not balanced in some way (or the cooled gas does not form stars)
then the stellar component of cD galaxies should be much more massive
than is observed, merely by the accretion of cooled intracluster gas.
By measuring the quantity of young stars and warm/cold gas in these
objects we can determine how well heating balances cooling.

The emission-line filaments are markers of feedback in massive
galaxies and to use them as such we need to understand their
composition, origin and lifetime.  Here we study two of the brightest
and most extensive emission-line nebulae around central cluster
galaxies NGC~1275 (at redshift $z=0.0176$, in the Perseus Cluster) and
NGC~4696 (redshift $z=0.009867$, in the Centaurus cluster) using the
Spitzer Infrared Spectrograph (IRS). The H$\alpha$ filaments around
NGC~1275, which extend well beyond the body of the galaxy, have warm
(2000K) molecular hydrogen associated with them. This motivates us to
search for much cooler (few $\times100$K) molecular hydrogen in these
regions using the rotational emission lines accessible to the IRS.

The central cD galaxy in the nearby Perseus cluster, NGC~1275,
is surrounded by spectacular H$\alpha$ filaments which
stretch over 100~kpc. These were first reported by \citet{Minkowski57}
and imaged by \citet{Lynds70}, \citet{McNamara96} and recently by
\citet{Conselice01}. The Perseus cluster is the brightest cluster of
galaxies (in flux units) at X-ray wavelengths with the emission peaking on NGC~1275
(\citealt{Fabian81, Fabian00, Fabian03a}). The nucleus of NGC~1275
also powers an FRI radio source, 3C~84 (\citealt{Pedlar90}).  Bubbles
of relativistic plasma have been blown by the jets from the nucleus
and have displaced the X-ray emitting intracluster gas
(\citealt{Boehringer93, Churazov01, Fabian00, Fabian02a, Fabian03a}).
It is likely that the H$\alpha$ filaments avoid the bubbles in three
dimensional space, although there is no obvious anti-correlation as
seen in projection on the sky; They appear to act like streamlines
showing that the flow behind the outer bubble, and thus the whole
inner medium, is not turbulent and may well have been dragged out from
the central region (\citealt{Fabian03b}, \citealt{Hatch06}). Molecular
hydrogen (\citealt{Krabbe00,Jaffe01,Edge02}) and CO
(\citealt{Salome06} and references therein) have been observed and
mapped in the inner regions for some time.

NGC~4696 at the centre of the Centaurus cluster has
H$\alpha$ filaments discovered by \citet{Fabian82} and a strong dust
lane (\citealt{Shobbrook66}). \citet{Sparks89} and \citet{Crawford05}
have mapped the nebulosity which appears to surround the dust lane.

The ionization state of the H$\alpha$ filaments in central cluster
galaxies is low, with [OI]$\lambda$6300, [OII]$\lambda3727$ and
[NII]$\lambda6584$ being prominent in optical spectra as well as
Balmer lines. Molecular H$_2$ is also common (e.g. \citealt{Jaffe97,
  Donahue00, Edge02}), even in the outer filaments
(\citealt{Hatch05}). \citet{Egami06} have recently found very
strong rotational H$_2$ lines from the brightest cluster galaxy in
Zw\,3146, using Spitzer IRS data. Unlike with our objects, the
emission is not spatially resolved due to the high redshift of the
object ($z=0.29$). The authors remark that this source has the most luminous
pure rotational H$_2$ lines and the largest mass of molecular hydrogen
known in a brightest cluster galaxy. A Spitzer IRS spectrum of the centre of
NGC~1275 is presented by \citet{Weedman05}. Forbidden atomic lines
and at least one molecular line are seen sitting on a large continuum
associated with the active nucleus. \citet{Kaneda05, Kaneda07} have
taken low resolution Spitzer IRS spectra of NGC~4696 and find PAH
features and [NeII] line emission.

What is ionizing and exciting the filamentary gas remains uncertain.
It is not simply an active nucleus (\citealt{Johnstone88, Sabra00}),
nor do a range of alternative mechansims provide a comprehensive
explanation (e.g. \citealt{Crawford92, Donahue00, Sabra00, Wilman02}).
The UV emission from massive young stars is likely to play an
important role (\citealt{Johnstone87, Allen95, Crawford99}), although
this may not be the case for the outer filaments. The lifetime of the
filaments is not known, although the immense size of the system around
NGC\,1275 ($80\kpc$) and the low velocity spread seen in the filaments
($\sim300\kmps$ or less) indicates that they may last over $10^8\yr$.

In this paper we present mid infra-red Spitzer spectra of off-nuclear extended
emission-line regions in NGC~1275 and NGC~4696. They probe a new region in
temperature space ($\sim 300-400$K) which lies between the very
cold regions observed in transitions of the CO molecule and the
warmer regions seen though ro-vibrational transitions of the hydrogen molecule.
These emission-line filaments seem to be a key marker of feedback in cooling
core clusters \citep{Hu85,Crawford99} even though they do not contain a significant fraction of the
total gas mass. It is by studying regions located away
from the nucleus of the galaxies that we expect to be able to make progress in
the understanding of the heating and ionization mechanisms of these
filaments as they offer a simpler environment than is found close to
the nucleii of the galaxies.

\section{Observations and data reduction}
\label{observations}

\subsection{Mid-infrared Spitzer Data}
\label{spitzer_data}
Spitzer IRS observations of off-nuclear regions in NGC~1275 and NGC~4696,
were made during two
observing campaigns in 2005 July and 2006 March (IRSX005200 and
IRSX006300).
For each object both
the Short Wavelength High Resolution (SH) and Long Wavelength High
Resolution (LH) spectrographs were used, giving an effective resolving
power of R$\sim600$ and a spectral coverage from $\sim10\mu$m to
$\sim37\mu$m. The entrance aperture size of the two
spectrographs is rather different, being 4.7$\times$11.3 arcsec for the SH
and 11.1$\times$22.3 arcsec for the LH spectrographs respectively.

Table \ref{tab:observations} lists details of the Spitzer observations.
Three off-nuclear regions were observed in NGC~1275, chosen to be
coincident with particular regions identified in the H$\alpha$+[NII] map of
\citet{Conselice01}, as well as a blank sky region.
In NGC~4696 we covered the brightest part of the H$\alpha$ nebula to
the South West of the nucleus.

\begin{figure}
\protect\resizebox{\columnwidth}{!}
{\includegraphics{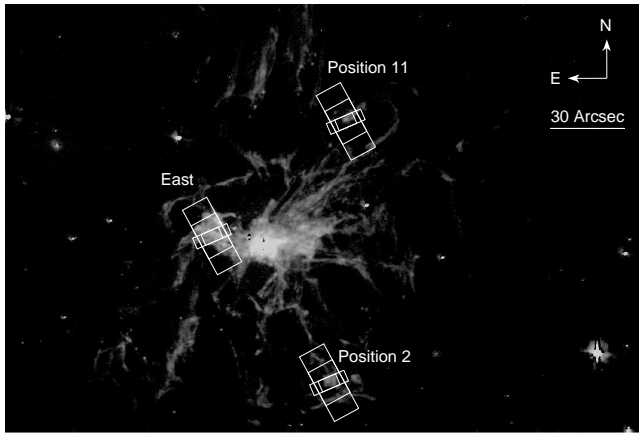}}
\caption{Spitzer spectrograph apertures overlaid on
an H$\alpha$+[NII] image of the central region of NGC~1275
\citep{Conselice01}. Small boxes show the positions of the SH
spectrograph aperture while the larger boxes show the positions of the
LH spectrograph aperture.}
\label{fig:permap}
\end{figure}

\begin{figure}
\protect\resizebox{\columnwidth}{!}
{\includegraphics[angle=180]{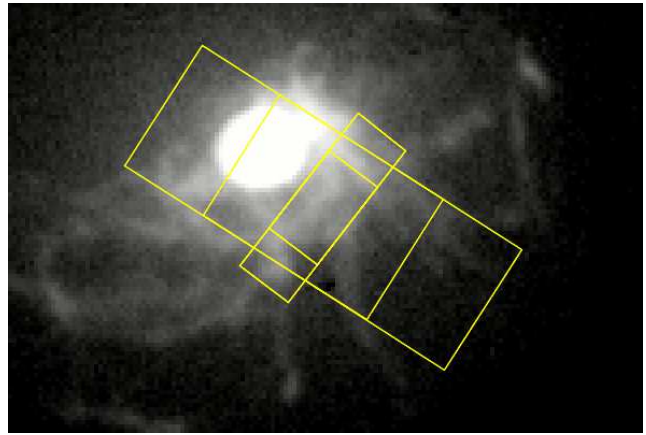}}
\caption{Spitzer spectrograph apertures overlaid on
an H$\alpha$+[NII] image of the central region of NGC~4696
\citep{Crawford05}. North is up and East to the left. Small boxes show
the positions of the SH spectrograph aperture while the larger boxes
show the positions of the LH spectrograph aperture.}
\label{fig:cenmap}
\end{figure}

Figures \ref{fig:permap} and \ref{fig:cenmap} show the outline of the
Spitzer spectrograph apertures
overlaid on H$\alpha$+[NII] images of the central cluster galaxies.
There are two slightly different pointings (exposure ids) for each
target position, these
being the standard nod positions in the Staring Mode Astronomical
Observation Template. These
positions locate the target at one third and two thirds of the way
along the spectrograph aperture. The SH spectrograph aperture positions
are shown by the smaller
rectangles whereas the LH spectrograph apertures are the larger
rectangles which are approximately orthogonal to the SH spectrograph
apertures. The specific pointing positions for each exposure id are
given in table \ref{tab:observations}.

\begin{table*}
\caption{Log of Spitzer Observations.}
\begin{center}
\begin{tabular}{lrrrrrrrr} \\ \hline

\multicolumn{1}{l}{Target} &
\multicolumn{1}{r}{Redshift} &
\multicolumn{1}{r}{Obs date} &
\multicolumn{1}{r}{AOR key} &
\multicolumn{1}{r}{Mode} &
\multicolumn{1}{r}{Exposure id} &
\multicolumn{1}{c}{RA} &
\multicolumn{1}{c}{Dec} &
\multicolumn{1}{r}{Exposure time} \\ 

\multicolumn{1}{l}{} &
\multicolumn{1}{r}{} &
\multicolumn{1}{r}{} &
\multicolumn{1}{r}{} &
\multicolumn{1}{r}{} &
\multicolumn{1}{r}{} &
\multicolumn{2}{c}{(2000)} &
\multicolumn{1}{r}{(seconds)} \\ \hline

\multicolumn{1}{l}{} &
\multicolumn{1}{r}{} &
\multicolumn{1}{r}{} &
\multicolumn{1}{r}{} &
\multicolumn{1}{r}{} \\

\multicolumn{1}{l}{NGC~1275 East} &
\multicolumn{1}{l}{0.0176} &
\multicolumn{1}{r}{2006 Mar 14} &
\multicolumn{1}{r}{r14536704} &
\multicolumn{1}{r}{SH} &
\multicolumn{1}{r}{0002} &
\multicolumn{1}{c}{03:19:49.9} &
\multicolumn{1}{c}{+41:30:48} &    
\multicolumn{1}{r}{481.69} \\

\multicolumn{1}{l}{} &
\multicolumn{1}{l}{} &
\multicolumn{1}{r}{} &
\multicolumn{1}{r}{} &
\multicolumn{1}{r}{} &
\multicolumn{1}{r}{0003} &
\multicolumn{1}{c}{03:19:50.2} &
\multicolumn{1}{c}{+41:30:46} &
\multicolumn{1}{r}{481.69} \\

\multicolumn{1}{l}{} &
\multicolumn{1}{l}{} &
\multicolumn{1}{r}{} &
\multicolumn{1}{r}{} &
\multicolumn{1}{r}{LH} &
\multicolumn{1}{r}{0004} &
\multicolumn{1}{c}{03:19:49.9} &
\multicolumn{1}{c}{+41:30:44} &
\multicolumn{1}{r}{241.83} \\

\multicolumn{1}{l}{} &
\multicolumn{1}{l}{} &
\multicolumn{1}{r}{} &
\multicolumn{1}{r}{} &
\multicolumn{1}{r}{} &
\multicolumn{1}{r}{0005} &
\multicolumn{1}{c}{03:19:50.2} &
\multicolumn{1}{c}{+41:30:50} &
\multicolumn{1}{r}{241.83} \\

\multicolumn{1}{l}{NGC~1275 Position 2} &
\multicolumn{1}{l}{0.0176} &
\multicolumn{1}{r}{2006 Mar 14} &
\multicolumn{1}{r}{r14536448} &
\multicolumn{1}{r}{SH} &
\multicolumn{1}{r}{0002} &
\multicolumn{1}{c}{03:19:45.7} &
\multicolumn{1}{c}{+41:29:49} &
\multicolumn{1}{r}{1445.07} \\

\multicolumn{1}{l}{} &
\multicolumn{1}{l}{} &
\multicolumn{1}{r}{} &
\multicolumn{1}{r}{} &
\multicolumn{1}{r}{} &
\multicolumn{1}{r}{0003} &
\multicolumn{1}{c}{03:19:46.0} &
\multicolumn{1}{c}{+41:29:47} &
\multicolumn{1}{r}{1445.07} \\

\multicolumn{1}{l}{} &
\multicolumn{1}{l}{} &
\multicolumn{1}{r}{} &
\multicolumn{1}{r}{} &
\multicolumn{1}{r}{LH} &
\multicolumn{1}{r}{0004} &
\multicolumn{1}{c}{03:19:45.7} &
\multicolumn{1}{c}{+41:29:45} &
\multicolumn{1}{r}{725.48} \\

\multicolumn{1}{l}{} &
\multicolumn{1}{l}{} &
\multicolumn{1}{r}{} &
\multicolumn{1}{r}{} &
\multicolumn{1}{r}{} &
\multicolumn{1}{r}{0005} &
\multicolumn{1}{c}{03:19:46.0} &
\multicolumn{1}{c}{+41:29:51} &
\multicolumn{1}{r}{725.48} \\

\multicolumn{1}{l}{NGC~1275 Position 11} &
\multicolumn{1}{l}{0.0176} &
\multicolumn{1}{r}{2006 Mar 14} &
\multicolumn{1}{r}{r14536192} &
\multicolumn{1}{r}{SH} &
\multicolumn{1}{r}{0002} &
\multicolumn{1}{c}{03:19:45.0} &
\multicolumn{1}{c}{+41:31:34} &
\multicolumn{1}{r}{1445.07} \\

\multicolumn{1}{l}{} &
\multicolumn{1}{l}{} &
\multicolumn{1}{r}{} &
\multicolumn{1}{r}{} &
\multicolumn{1}{r}{} &
\multicolumn{1}{r}{0003} &
\multicolumn{1}{c}{03:19:45.3} &
\multicolumn{1}{c}{+41:31:32} &
\multicolumn{1}{r}{1445.07} \\

\multicolumn{1}{l}{} &
\multicolumn{1}{l}{} &
\multicolumn{1}{r}{} &
\multicolumn{1}{r}{} &
\multicolumn{1}{r}{LH} &
\multicolumn{1}{r}{0004} &
\multicolumn{1}{c}{03:19:45.0} &
\multicolumn{1}{c}{+41:31:30} &
\multicolumn{1}{r}{725.48} \\

\multicolumn{1}{l}{} &
\multicolumn{1}{l}{} &
\multicolumn{1}{r}{} &
\multicolumn{1}{r}{} &
\multicolumn{1}{r}{} &
\multicolumn{1}{r}{0005} &
\multicolumn{1}{c}{03:19:45.3} &
\multicolumn{1}{c}{+41:31:36} &
\multicolumn{1}{r}{725.48} \\

\multicolumn{1}{l}{NGC~1275 Blank} &
\multicolumn{1}{l}{0.0176} &
\multicolumn{1}{r}{2006 Mar 14} &
\multicolumn{1}{r}{r16754176} &
\multicolumn{1}{r}{SH} &
\multicolumn{1}{r}{0000} &
\multicolumn{1}{c}{03:19:53.6} &
\multicolumn{1}{c}{+41:28:35} &
\multicolumn{1}{r}{963.38} \\

\multicolumn{1}{l}{} &
\multicolumn{1}{l}{} &
\multicolumn{1}{r}{} &
\multicolumn{1}{r}{} &
\multicolumn{1}{r}{} &
\multicolumn{1}{r}{0001} &
\multicolumn{1}{c}{03:19:53.9} &
\multicolumn{1}{c}{+41:28:33} &
\multicolumn{1}{r}{963.38} \\

\multicolumn{1}{l}{} &
\multicolumn{1}{l}{} &
\multicolumn{1}{r}{} &
\multicolumn{1}{r}{} &
\multicolumn{1}{r}{LH} &
\multicolumn{1}{r}{0002} &
\multicolumn{1}{c}{03:19:53.6} &
\multicolumn{1}{c}{+41:28:31} &
\multicolumn{1}{r}{483.65} \\

\multicolumn{1}{l}{} &
\multicolumn{1}{l}{} &
\multicolumn{1}{r}{} &
\multicolumn{1}{r}{} &
\multicolumn{1}{r}{} &
\multicolumn{1}{r}{0003} &
\multicolumn{1}{c}{03:19:53.9} &
\multicolumn{1}{c}{+41:28:37} &
\multicolumn{1}{r}{483.65} \\

\multicolumn{1}{l}{NGC~4696} &
\multicolumn{1}{l}{0.009867} &
\multicolumn{1}{r}{2005 Jul 9} &
\multicolumn{1}{r}{r14537216} &
\multicolumn{1}{r}{SH} &
\multicolumn{1}{r}{0002} &
\multicolumn{1}{c}{12:48:48.7} &
\multicolumn{1}{c}{-41:18:42} &
\multicolumn{1}{r}{2408.44} \\

\multicolumn{1}{l}{} &
\multicolumn{1}{l}{} &
\multicolumn{1}{r}{} &
\multicolumn{1}{r}{} &
\multicolumn{1}{r}{} &
\multicolumn{1}{r}{0003} &
\multicolumn{1}{c}{12:48:48.9} &
\multicolumn{1}{c}{-41:18:45} &
\multicolumn{1}{r}{2408.44} \\

\multicolumn{1}{l}{} &
\multicolumn{1}{l}{} &
\multicolumn{1}{r}{} &
\multicolumn{1}{r}{} &
\multicolumn{1}{r}{LH} &
\multicolumn{1}{r}{0004} &
\multicolumn{1}{c}{12:48:48.5} &
\multicolumn{1}{c}{-41:18:45} &
\multicolumn{1}{r}{725.48} \\

\multicolumn{1}{l}{} &
\multicolumn{1}{l}{} &
\multicolumn{1}{r}{} &
\multicolumn{1}{r}{} &
\multicolumn{1}{r}{} &
\multicolumn{1}{r}{0005} &
\multicolumn{1}{c}{12:48:49.1} &
\multicolumn{1}{c}{-41:18:42} &
\multicolumn{1}{r}{725.48} \\





\hline
\end{tabular}\\
\label{tab:observations}
\end{center}
\end{table*}

The observations were processsed by the Spitzer Science Center through
pipeline version S13.2.0. We use the `bcd.fits', `func.fits' and
`bmask.fits' files as the starting point for our reduction.

Individual Data Collection Event (DCE) files within each exposure id
were first averaged together and an array of external uncertainties
corresponding to the standard error on the mean was calculated. A new
bmask file was also generated, this being the logical `or' of all the
bmasks for the input DCEs. The NGC~1275 East pointing has only one DCE
in each spectrograph so in these cases the pipline based
uncertainties, which are computed from the ramp fitting of the
multiple detector readouts, are used for the uncertainty arrays.

The NGC~1275 pointings were then background subtracted using the
NGC~1275 blank field pointing, propagating the uncertainty arrays in
quadrature and again creating a new bmask file that is the logical
`or' of the input masks.

In the case of the NGC~4696 pointing we do not have a concurrent
sky-background observation. For the LH data, we used an average of
background observations (scaled to the expected background level at
the position of the NGC~4696 pointing using values from the Spitzer
SPOT software) taken for standard star calibrations in the same
observing campaign to effect sky subtraction. However, for the SH
data, the analagous background observations were too short compared
with the NGC~4696 data and added considerable noise when subtracted.
We therefore leave these as not sky subtracted. One consequence of not
having a concurrent sky background observation is that the rogue
pixels in the detectors are not well corrected (or corrected at all in
the case of the SH data). We therefore used the IRSclean\_mask
software from the Spitzer contributed software
site\footnote{\url{http://ssc.spitzer.caltech.edu/archanaly/contributed/browse.html}}
to interpolate over the rogue pixels for these observations.

The data from the SH and LH spectrographs are not rectilinear in
either the spatial or spectral direction and therefore require
software which has a knowledge of the detailed distortions in order to
be able to extract one-dimesional spectra. We used the Spitzer Spice
software (version 1.3beta1 and version 1.4) for this, extracting all
the spectral orders over the full aperture in both spectrographs. The
final step in the Spice software extraction is a `tuning' that applies
a flux calibration to the data. There are two possibilities at this
stage because the size of the point-spread function of the Spitzer
telescope is of the same order as the entrance aperture size. One
option applies an extended source calibration in which it is assumed
that the aperture is illuminated by a uniform surface brightness
source; as much flux is scattered into the aperture as is scattered
out of it. The other option is to apply the point-source calibration
which corrects for the fraction of flux from a well-centered
point-source that is lost from the aperture, as a function of
wavelength.

To determine whether the point-source or extended source calibration
is more appropriate for our data we have taken the continuum
subtracted H$\alpha$+[NII] maps and used them as proxies for the spatial
distribution of the infra-red line emission since we do not know its
spatial distribution.  Assuming that the H$\alpha$+[NII] emission is
distributed spatially like the infra-red emission allows us to make a
quantitative (but approximate) assessment of whether the emission,
which appears very clearly extended on the resolution of the optical
data, is actually extended at the resolution of the Spitzer
observations. We have measured the ratio of the flux in the Spitzer
apertures in the raw image and in the image smoothed using a gaussian
kernel of full-width half-maximum of 3.1 or 6.17 arcsec. These widths
are an approximation to the size of the Spitzer point spread function
at 15$\mu$m and 30$\mu$m (approximately the mid-points of the SH and
LH spectra respectively) as measured by fitting a one-dimensional
gaussian model to the azimuthal average of the central peak of a point
source generated by the Spitzer Stinytim point-spread function
modelling software. Table \ref{tab:calratio} lists these ratios for
each of the exposure ids.

\begin{table}
\caption[]{Ratio of H$\alpha$+[NII] fluxes in the Spitzer apertures measured from the
  original maps of \cite{Conselice01} (NGC~1275) and \cite{Crawford05}
  (NGC~4696) to those smoothed with a gaussian kernel of
  FWHM of 3.10 arcsec (for the SH spectra) and a FWHM of 6.17
  arcsec (for the LH spectra).}
\begin{tabular}{lrrr}
\hline
Region& Exposure id & Spectrograph & Flux ratio\\
\hline

NGC~1275 East          & 0002  & SH   & 1.00  \\
                      & 0003  & SH   & 1.06  \\
                      & 0004  & LH   & 1.10  \\
                      & 0005  & LH   & 1.11  \\
NGC~1275 Position 2    & 0002  & SH   & 1.07  \\
                      & 0003  & SH   & 1.13  \\
                      & 0004  & LH   & 1.04  \\
                      & 0005  & LH   & 1.08  \\
NGC~1275 Position 11   & 0002  & SH   & 1.09  \\
                      & 0003  & SH   & 1.11  \\
                      & 0004  & LH   & 1.07  \\
                      & 0005  & LH   & 1.01  \\
Centaurus             & 0002  & SH   & 0.99  \\
                      & 0003  & SH   & 1.00  \\
                      & 0004  & LH   & 1.16  \\
                      & 0005  & LH   & 1.06  \\
\hline

\end{tabular}
\label{tab:calratio}
\end{table}

Examination of Table \ref{tab:calratio} shows that there is up to 16 per cent
more H$\alpha$+[NII] flux measured in the Spitzer apertures in the
unsmoothed images compared with the smoothed images. The inverse of
the Spitzer Slitloss Correction Function\footnote{\url{http://ssc.spitzer.caltech.edu/irs/calib/extended_sources/index.html}}
shows that for a point source observed by Spitzer the factor
would be 1.63 and 1.52 at 15$\mu$m and 30$\mu$m in the SH and LH
spectrographs respectively, whereas a uniform brightness extended
source would give factors of 1.0 for both wavelengths. This indicates
that the data approximate much better to an extended source than a
point source. We therefore proceed using the extended source tuning
option in Spice.

\begin{figure}
\protect\resizebox{\columnwidth}{!}
{
\includegraphics[angle=270]{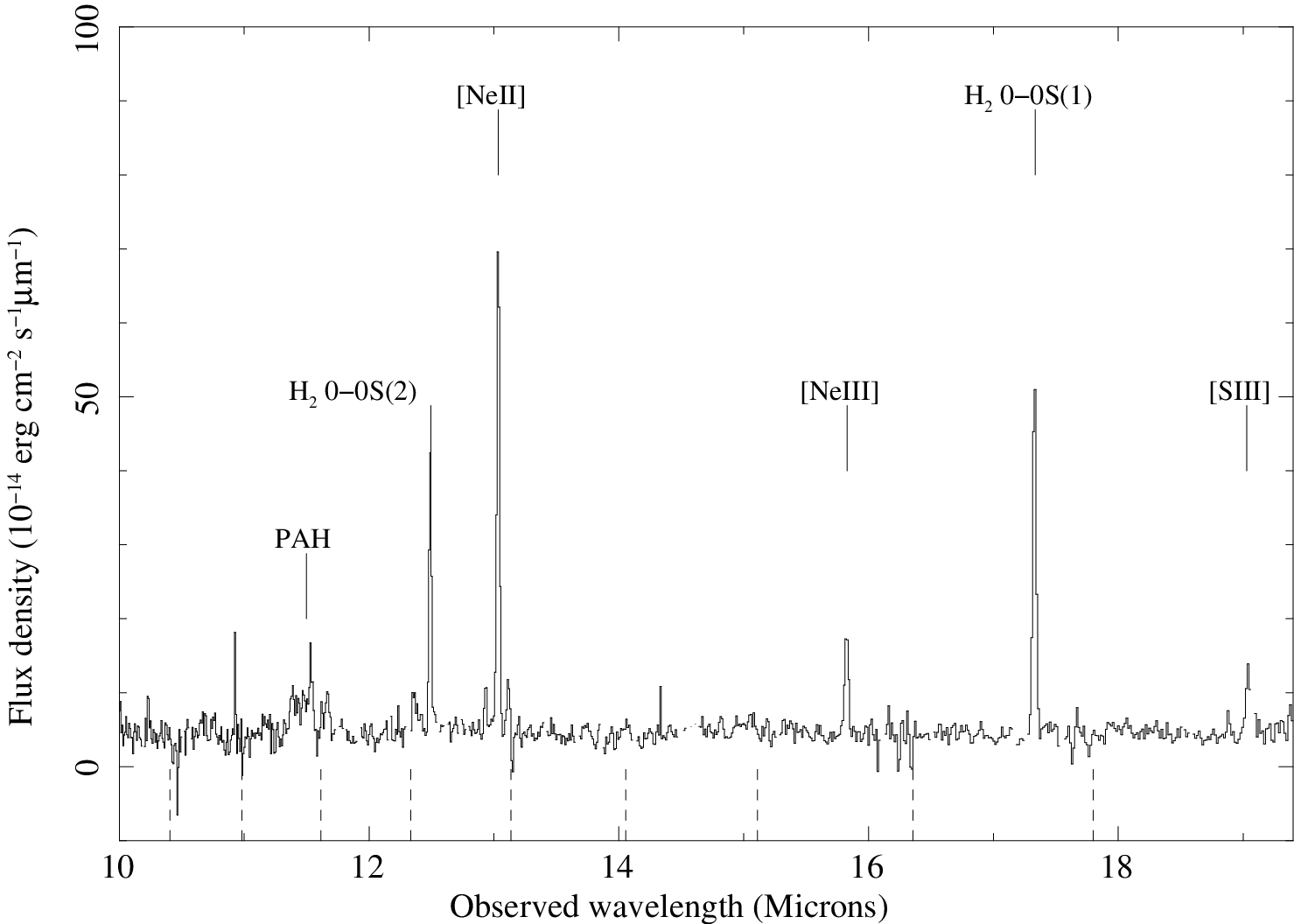}
}
\protect\resizebox{\columnwidth}{!}
{
\includegraphics[angle=270]{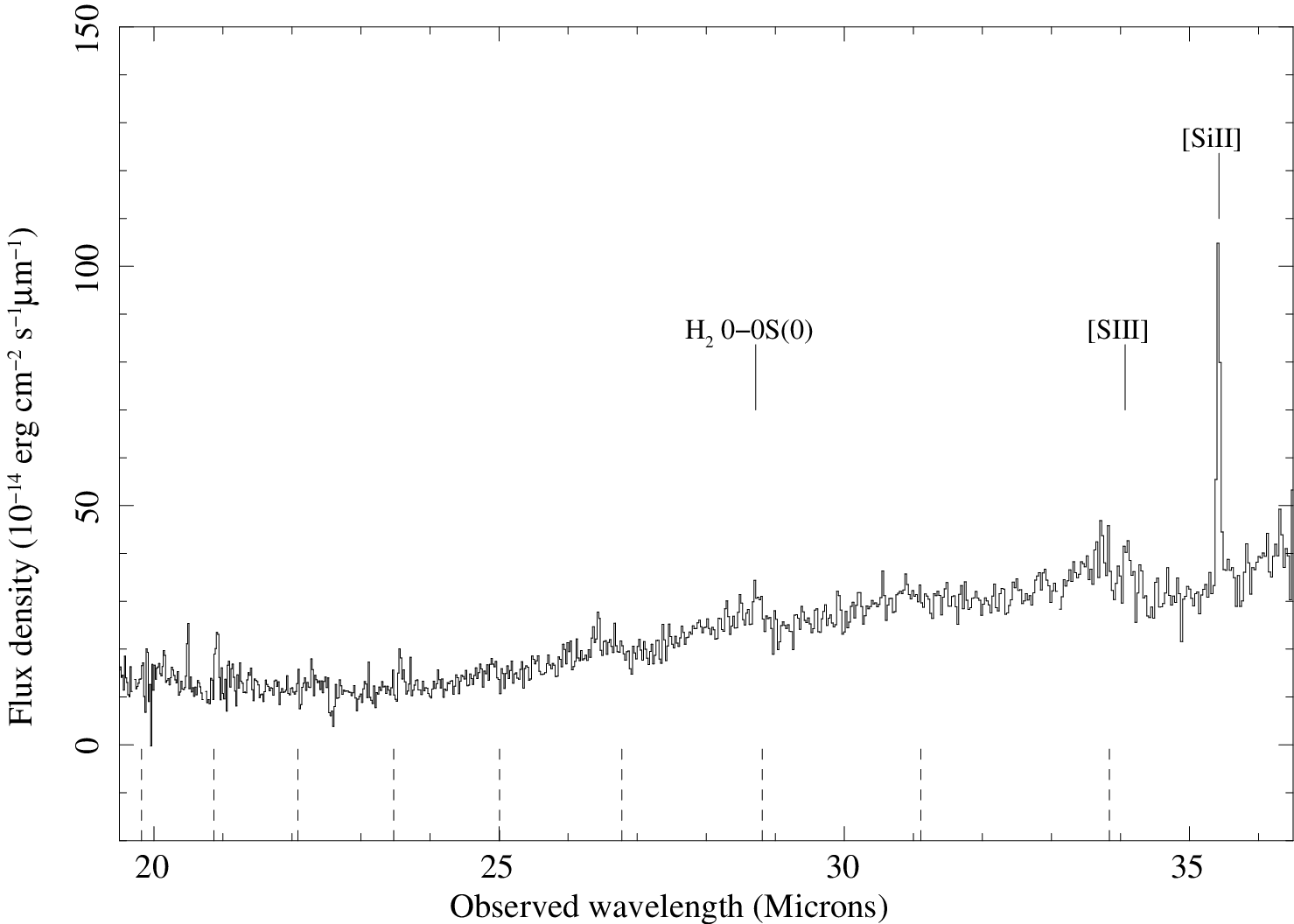}
}
\caption{Upper: Spitzer SH (exposure id 0002) and Lower: Spitzer LH
  (exposure id 0004) spectra for the NGC~1275 East pointing. Dashed
  vertical lines show the boundary between adjacent spectral orders.}
\label{fig:pereastspectra}
\end{figure}

\begin{figure}
\protect\resizebox{\columnwidth}{!}
{
\includegraphics[angle=270]{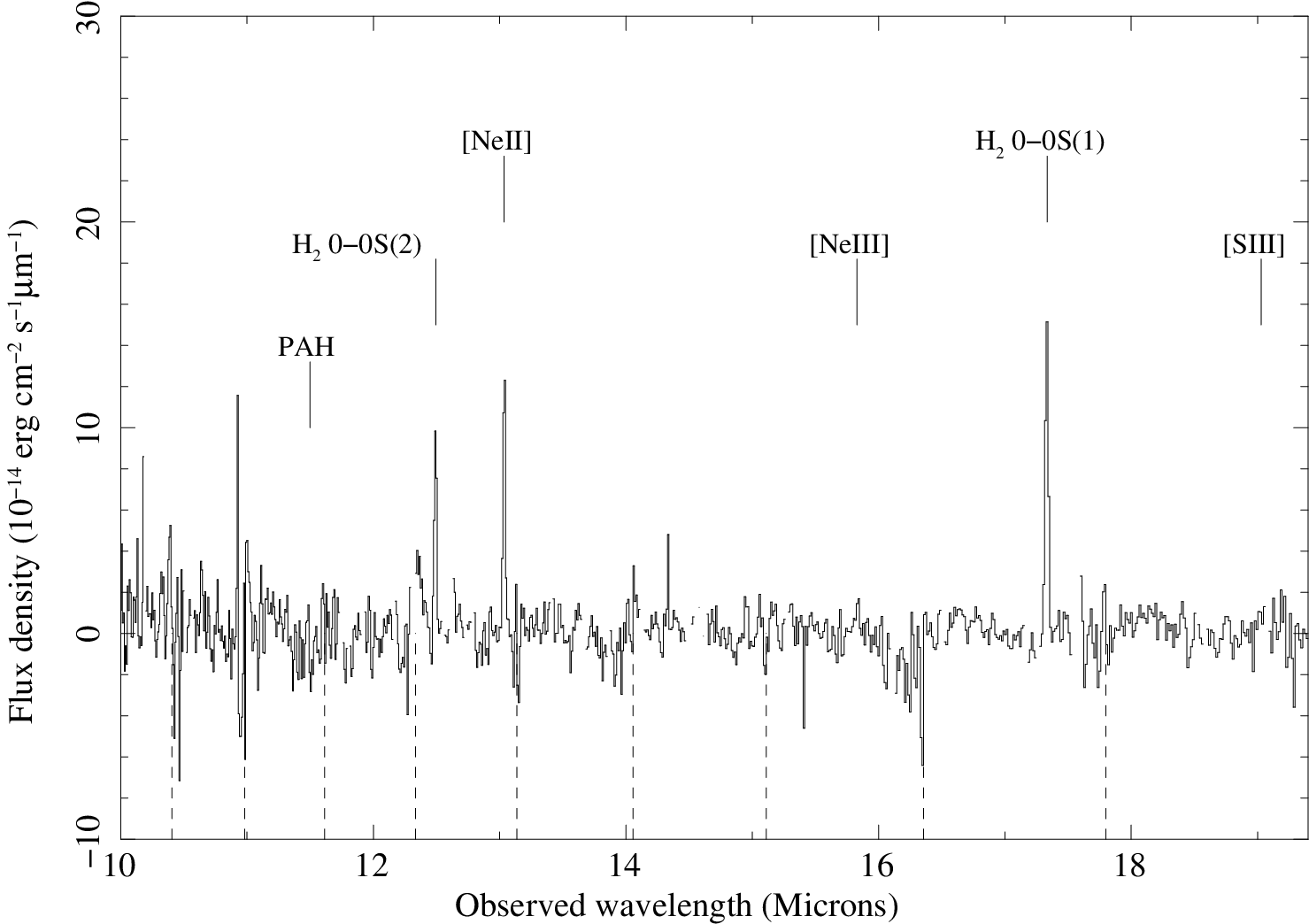}
}
\protect\resizebox{\columnwidth}{!}
{
\includegraphics[angle=270]{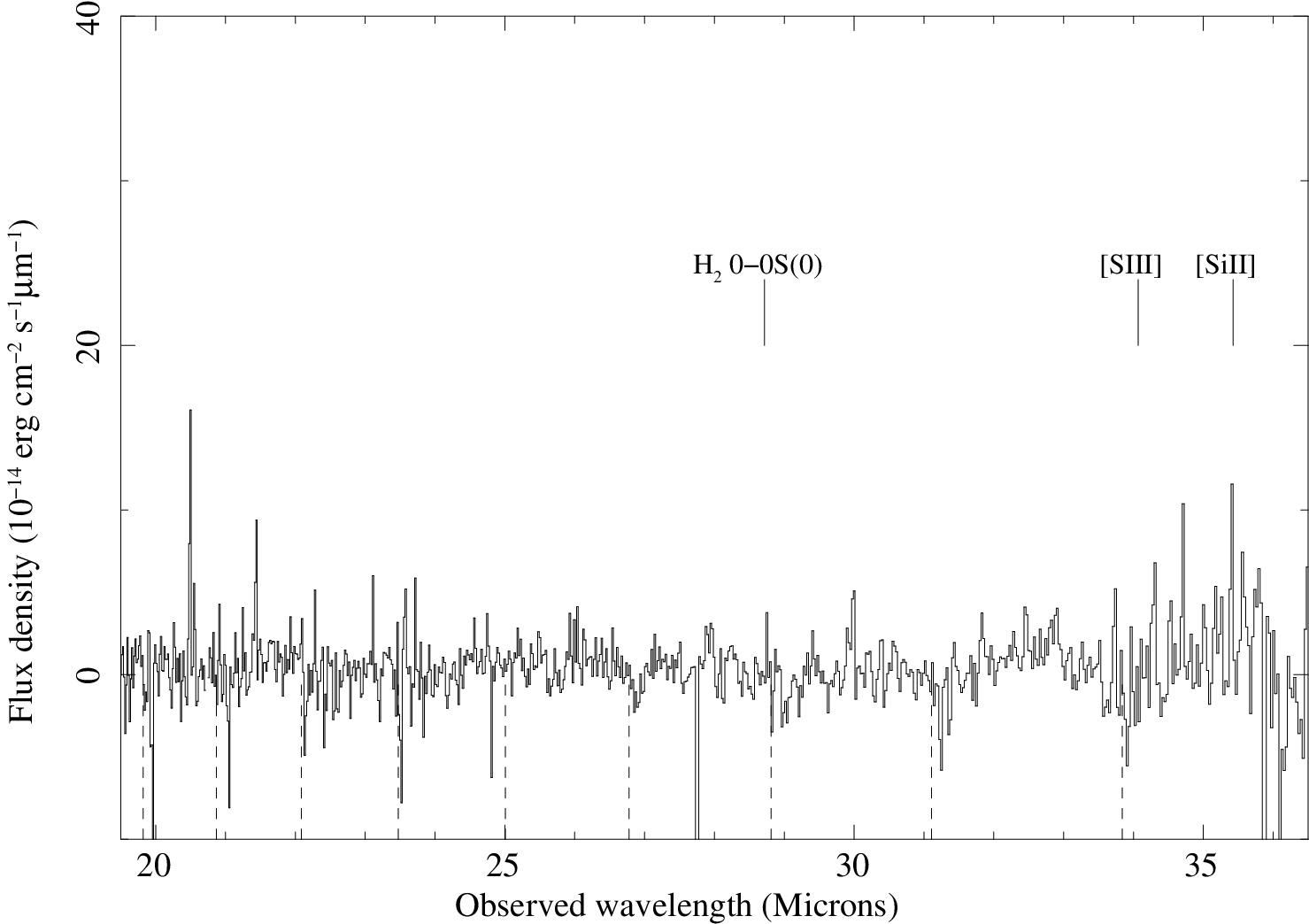}
}
\caption{Upper: Spitzer SH (exposure id 0002) and Lower: Spitzer LH
  (exposure id 0004) spectra for the NGC~1275 Position 2. Dashed
  vertical lines show the boundary between adjacent spectral orders.}
\label{fig:per2spectra}
\end{figure}

\begin{figure}
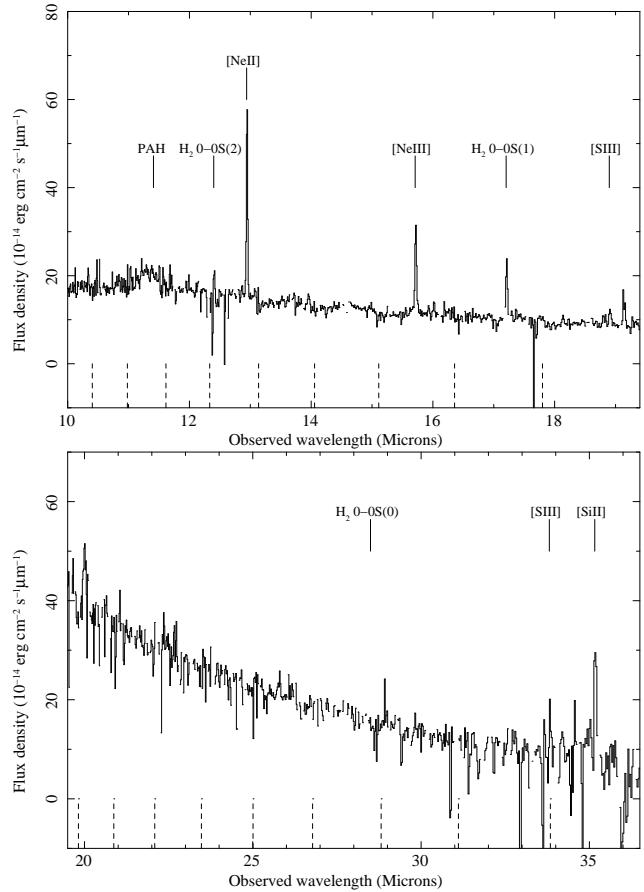

\protect\resizebox{\columnwidth}{!}
{
\includegraphics[angle=270]{cen_expid2_avg_bcd_clean.extsrc_spect2.ps}
}

\protect\resizebox{\columnwidth}{!}
{
\includegraphics[angle=270]{cen_expid4_avg_skysub_bcd_clean.extsrc_spect2_pretty.ps}
}

\caption{Upper: Spitzer SH (exposure id 0002) and Lower: Spitzer LH
  (exposure id 0004) spectra for the NGC~4696 pointing. Dashed
  vertical lines show the boundary between adjacent spectral
  orders. Note that the SH data are not background subtracted.}
\label{fig:censpectra}
\end{figure}

The final part of the reduction involves merging the individual
spectral orders into one spectrum and converting the flux units to per
unit wavelength interval. The sensitivity of the spectrograph drops
significantly at the end of each order leading to an increase in the
noise, but because there is an overlap in the wavelength coverage of
adjacent orders (and the wavelength binning is the same at
corresponding wavelengths) we are able to use the following
prescription to clean the final spectrum: In the order overlap regions
we choose data from the lower order number spectrum in preference to
that of the higher order number, except that we discard the 6 shortest
wavelength bins in the highest order spectrum and the 4 shortest
wavelength bins of all other orders. The 6 longest wavelength bins in
the spectrum are also discarded. The spectral binning used by the
Spice extraction routine varies as a function of wavelength to preseve
a constant velocity resolution of approximately 240\kmps per bin. We
show examples of the final spectra in Figs. \ref{fig:pereastspectra},
\ref{fig:per2spectra}, \ref{fig:per11spectra} and
\ref{fig:censpectra}, with the expected positions of some emission
lines marked. The dashed vertical lines show the boundary of adjacent
spectral orders.

The emission lines in the spectra were fitted individually using a
gaussian model for the line plus a straight line to account for the
local continuum. The fitting was done using the QDP
program\footnote{\url{http://wwwastro.msfc.nasa.gov/qdp}} to minimize
the chi-square statistic between the model and the data, subject to
the propagated uncertainty in the flux at each wavelength bin.  Line
parameters and uncertainties are given in Tables
\ref{tab:pereastfluxes}, \ref{tab:per2fluxes}, \ref{tab:per11fluxes},
and \ref{tab:cenfluxes}. The uncertainties were generally calculated
(for all pointings except the NGC~1275 East region) from the
$\Delta\chi^2=1.0$ criterion (\citealt{Lampton76}). This corresponds
to a 68 per cent confidence region or a 1$\sigma$ (gaussian
equivalent) confidence region for one interesting parameter. In the
cases where lines are not detected we set a 3$\sigma$ upper limit on
the line flux by fixing the position and width of the line and
increasing the flux until the value of chi-square increases by 9 from
the value obtained with no line present. Where no uncertainty on a
parameter is listed that parameter was fixed. Line widths for detected
lines were left as free parameters in the fitting. However, except for
the PAH 11.3$\mu$m line the widths are not resolved beyond the
instrumental resolution and are therefore not listed.

The values of the minimum chi-square that were
obtained when fitting the NGC~4696 short wavelength spectra were
significantly above the number of degrees of freedom, whereas the values of the minimum chi-square that were
obtained when fitting the NGC~1275 East spectra were
significantly below the number of degrees of freedom. In both cases
this indicates that 
the uncertainties on each bin are wrong. In the case of the NGC~4696
short wavelength data there is additional power in the spectra not
accounted for in the uncertainty arrays which may be due to the
presence of fringes
(http://ssc.spitzer.caltech.edu/postbcd/irsfringe.html) that have not
been completely removed in the pipleline sofware. We tried to correct
for these using the irsfringe tool but were unsuccessful. In the case
of the NGC~1275 East spectra there is only one DCE and the
uncertainties are propagated from the ramp fitting. This leads to
uncertainties which are too big.

Since using the
$\Delta\chi^2=1.0$ criterion under either of these conditions leads to
uncertainties on the line parameters which are either too large or too
small we have
adopted the `Ratio of Variances' technique (\citealt{Lampton76}) to
rescale the critical value of delta chi-square used to calculate the
line parameter uncertainties. We note in passing that the spectral
extraction method used in the Spice software does introduce
correlations between adjacent spectral bins but the Spitzer Science
Centre advises that the degreee of correlation is small because the
dispersion direction in the spectral images is almost parallel with
the pixel direction.

\begin{table*}
\caption[]{Properties of emission lines in the NGC~1275 East region
  Spitzer spectra. Note that the line widths are the gaussian sigma
  values.}
\begin{tabular}{lrrrrrrrrrr}
\hline
Exposure id & Line & Wavelength & Width ($\sigma$) & Flux \\
                     &               & (microns)           & (microns) & ($\times 10^{-14}\ergpscmps$) \\ \hline

0002 & PAH 11.3             & 11.51$\pm$0.01     & 0.11$\pm$0.01     & 1.43$\pm$0.16 \\
     & H$_{2}$ 0-0S(2) 12.28 & 12.4899$\pm$0.0003 & -                 & 0.85$\pm$0.03 \\
     & [NeII] 12.81          & 13.0342$\pm$0.0002 & -                 & 1.78$\pm$0.03 \\
     & [NeIII] 15.56         & 15.8241$\pm$0.0007 & -                 & 0.54$\pm$0.03 \\
     & H$_{2}$ 0-0S(1) 17.04 & 17.3280$\pm$0.0009 & -                 & 1.79$\pm$0.09 \\
     & [SIII] 18.71          & 19.040$\pm$0.003   & -                 & 0.39$\pm$0.07 \\

0003 & PAH 11.3             & 11.56$\pm$0.03     & 0.14$\pm$0.03     & 1.15$\pm$0.30 \\
     & H$_{2}$ 0-0S(2) 12.28 & 12.4906$\pm$0.0003 & -                 & 0.87$\pm$0.03 \\
     & [NeII] 12.81    & 13.0350$\pm$0.0003       & -                 & 1.51$\pm$0.04 \\
     & [NeIII] 15.56   & 15.8236$\pm$0.0010       & -                 & 0.44$\pm$0.03 \\
     & H$_{2}$ 0-0S(1) 17.04 & 17.3270$\pm$0.0010 & -                  & 1.98$\pm$0.12 \\
     & [SIII] 18.71    & 19.040$\pm$0.003         & -                 & 0.16$\pm$0.04 \\

0004 & H$_{2}$ 0-0S(0) 28.22 & 28.72              & -                 & $<$0.73   \\
     & [SIII] 33.48    & 34.07                    & -                 & $<$1.68 \\
     & [SiII] 34.82    & 35.421$\pm$0.001         & -                 & 4.46$\pm$0.18 \\

0005 & H$_{2}$ 0-0S(0) 28.22& 28.72               & -             & $<$0.75   \\
     & [SIII] 33.48    & 34.07                    & -             & $<$0.75\\
     & [SiII] 34.82    & 35.415$\pm$0.001         & -             & 3.59$\pm$0.17 \\

\hline
\end{tabular}
\label{tab:pereastfluxes}
\end{table*}

\begin{table*}
\caption[]{Properties of emission lines in the NGC~1275 Position 2
  region Spitzer spectra. Note that the line widths are the gaussian sigma
  values.}
\begin{tabular}{lrrrrrrrrrr}
\hline
Exposure id & Line & Wavelength & Width ($\sigma$)& Flux \\
                     &               & (microns)           & (microns) & ($\times 10^{-14}\ergpscmps$) \\ \hline
0002 & PAH 11.3              & 11.51              & 0.11     &  $<0.11$\\
     & H$_{2}$ 0-0S(2) 12.28 & 12.4920$\pm$0.0008 & -         & 0.24$\pm$0.02 \\
     & [NeII] 12.81          & 13.0358$\pm$0.0006 & -         & 0.30$\pm$0.02 \\
     & [NeIII] 15.56         & 15.8241            & -         & $<$0.19 \\
     & H$_{2}$ 0-0S(1) 17.04 & 17.3317$\pm$0.0004 &            & 0.47$\pm$0.02 \\
     & [SIII] 18.71          & 19.040             & -         & $<$0.10 \\

0003 & PAH 11.3              & 11.51              & 0.11       &  $< 0.11$\\
     & H$_{2}$ 0-0S(2) 12.28 & 12.4925$\pm$0.0010 & -           & 0.27$\pm$0.03 \\
     & [NeII] 12.81          & 13.0382$\pm$0.0011 & -          & 0.21$\pm$0.03 \\
     & [NeIII] 15.56         & 15.8241            & -          & $<0.11$ \\
     & H$_{2}$ 0-0S(1) 17.04 & 17.3334$\pm$0.0006 & -          & 0.42$\pm$0.02 \\
     & [SIII] 18.71          & 19.04              & -          & $<$0.07 \\

0004 & H$_{2}$ 0-0S(0) 28.22 & 28.72              & -           & $<$0.15 \\
     & [SIII] 33.48          & 34.07             & -            & $<$0.27 \\
     & [SiII] 34.82          & 35.43             & -            & $<$0.63 \\

0005 & H$_{2}$ 0-0S(0) 28.22 & 28.72             & -            & $<$0.18 \\
     & [SIII] 33.48          & 34.07             & -            & $<$0.36 \\
     & [SiII] 34.82          & 35.43             & -            & $<$0.69 \\
\hline

\end{tabular}
\label{tab:per2fluxes}
\end{table*}

\begin{table*}
\caption[]{Properties of emission lines in the NGC~1275 Position 11
  region Spitzer spectra. Note that the line widths are the gaussian sigma
  values.}
\begin{tabular}{lrrrrrrrrrr}

\hline
Exposure id & Line & Wavelength & Width ($\sigma$)& Flux \\
            &               & (microns)           & (microns) & ($\times 10^{-14}\ergpscmps$) \\ \hline

0002 & PAH 11.3              & 11.51              & 0.11     &  $<0.27$\\
     & H$_{2}$ 0-0S(2) 12.28 & 12.4955$\pm$0.0013 & -         & 0.13$\pm$0.02 \\
     & [NeII] 12.81          & 13.0402$\pm$0.0008 & -        & 0.29$\pm$0.03 \\
     & [NeIII] 15.56         & 15.8241            & -        & $<$0.07 \\
     & H$_{2}$ 0-0S(1) 17.04 & 17.3342$\pm$0.0008 & -         & 0.25$\pm$0.02 \\
     & [SIII] 18.71          & 19.040             & -        & $<$0.16 \\

0003 & PAH 11.3              & 11.51              & 0.11     &  $< 0.19$\\
     & H$_{2}$ 0-0S(2) 12.28 & 12.4958$\pm$0.0012 & -         & 0.13$\pm$0.02 \\
     & [NeII] 12.81          & 13.0411$\pm$0.0011 & -        & 0.15$\pm$0.02 \\
     & [NeIII] 15.56         & 15.8241            & -        & $<0.08$ \\
     & H$_{2}$ 0-0S(1) 17.04 & 17.3371$\pm$0.012  & -         &  0.21$\pm$0.02 \\
     & [SIII] 18.71          & 19.04              & -        & $<$0.11 \\

0004 & H$_{2}$ 0-0S(0) 28.22 & 28.72              & -         & $<$0.07 \\
     & [SIII] 33.48          & 34.07              & -        & $<$0.38 \\
     & [SiII] 34.82          & 35.43              & -         & $<$0.49 \\

0005 & H$_{2}$ 0-0S(0) 28.22 & 28.72              & -         & $<$0.14 \\
     & [SIII] 33.48          & 34.07             & -          & $<$0.24 \\
     & [SiII] 34.82          & 35.43             & -          & $<$0.20 \\

\hline

\end{tabular}
\label{tab:per11fluxes}
\end{table*}

\begin{table*}
\caption[]{Properties of emission lines in the NGC~4696 Spitzer
  spectra. Note that the line widths are the gaussian sigma
  values.}
\begin{tabular}{lrrrrrrrrrr}
\hline
Exposure id & Line & Wavelength & Width ($\sigma$)& Flux\\
            &               & (microns)           & (microns) & ($\times 10^{-14}\ergpscmps$) \\ \hline

0002 & PAH 11.3              & 11.32$\pm$0.017    & 0.15$\pm$0.02    & 1.44$\pm$0.2   \\
     & H$_{2}$ 0-0S(2) 12.28 & 12.408$\pm$0.003   & -                 & 0.17$\pm$0.03  \\
     & [NeII] 12.81          & 12.9469$\pm$0.0003 & -                 & 1.13$\pm$0.03 \\
     & [NeIII] 15.56         & 15.7184$\pm$0.0005 & -                 & 0.66$\pm$0.02 \\
     & H$_{2}$ 0-0S(1) 17.04 & 17.2122$\pm$0.0011 & -                  & 0.45$\pm$0.03 \\
     & [SIII] 18.71          & 18.9110$\pm$0.004  & -                  & 0.14$\pm$0.03 \\

0003 & PAH 11.3              & 11.33$\pm$0.016    & 0.11$\pm$0.019    & 1.75$\pm$0.21\\
     & H$_{2}$ 0-0S(2) 12.28 & 12.411$\pm$0.004   & -                  & 0.18$\pm$0.04  \\
     & [NeII] 12.81          & 12.9488$\pm$0.0003 & -                  & 0.98$\pm$0.03 \\
     & [NeIII] 15.56         & 15.7207$\pm$0.0004 & -                  & 0.58$\pm$0.02 \\
     & H$_{2}$ 0-0S(1) 17.04 & 17.2142$\pm$0.0006 & -                 & 0.42$\pm$0.02 \\
     & [SIII] 18.76          & 18.9133$\pm$0.005  & -                 & 0.13$\pm$0.04 \\

0004 & H$_{2}$ 0-0S(0) 28.22 & 28.497              &  -                 & $<$0.12 \\
     & [SIII] 33.48          & 33.811              & -                  & $<$0.94 \\
     & [SiII] 34.82          & 35.1808$\pm$0.0030  & -                & 1.81$\pm$0.15 \\

0005 & H$_{2}$ 0-0S(0) 28.22 & 28.497              & -                 & $<$0.10 \\
     & [SIII] 33.48          & 33.811              & -                 & $<$1.60 \\
     & [SiII] 34.82          & 35.1740$\pm$0.0021  & -                 & 3.05$\pm$0.18 \\

\hline

\end{tabular}
\label{tab:cenfluxes}
\end{table*}

\begin{table*}
\caption[]{H$\alpha$+[NII] emission-line fluxes in Spitzer apertures measured from
  maps of  \citet{Conselice01} (NGC~1275) and \citet{Crawford05}
  (NGC~4696). To account for the Spitzer point spread function, the
  maps were smoothed with a gaussian kernel of
  FWHM of 3.10 arcsec for the short wavelength aperture measurements
  and by gaussian with a FWHM of 6.17 arcsec for the long wavelength
  aperture measurements. No measurement is available for the NGC~4696
  LH apertures as these cover a region of the image in which the
  off-line image is saturated. The NGC~1275 values have been corrected
for a reddening corresponding to $A_R=0.4$.}
\begin{tabular}{lrrrrrrrrrr}
\hline
Region & Exposure id & Spectrograph & Flux \\
       &             &              & ($\times 10^{-13}\ergpscmps$) \\
\hline

NGC~1275 East         & 0002  & SH   & 5.8  \\
                      & 0003  & SH   & 4.8  \\
                      & 0004  & LH   & 16   \\
                      & 0005  & LH   & 15   \\
NGC~1275 Position 2   & 0002  & SH   & 1.1  \\
                      & 0003  & SH   & 1.0  \\
                      & 0004  & LH   & 2.7  \\
                      & 0005  & LH   & 2.8  \\
NGC~1275 Position 11  & 0002  & SH   & 0.83 \\
                      & 0003  & SH   & 0.68 \\
                      & 0004  & LH   & 1.7  \\
                      & 0005  & LH   & 1.4  \\
NGC~4696              & 0002  & SH   & 1.7  \\
                      & 0003  & SH   & 1.5  \\
                      & 0004  & LH   & --   \\
                      & 0005  & LH   & --   \\
\hline

\end{tabular}
\label{tab:hafluxes}
\end{table*}

\begin{figure}
\protect\resizebox{\columnwidth}{!}
{
\includegraphics[angle=270]{pos11_expid2_avg_skysub_extsrc_spect2.ps}
\par
}
\protect\resizebox{\columnwidth}{!}
{
\includegraphics[angle=270]{pos11_expid4_avg_skysub_bcd_clean.extsrc_spect2.ps}
}
\caption{Upper: Spitzer SH (exposure id 0002) and Lower: Spitzer LH
  (exposure id 0004) spectra for the NGC~1275 position 11. Dashed
  vertical lines show the boundary between adjacent spectral orders.}
\label{fig:per11spectra}
\end{figure}

\subsection{Supporting Near-infrared Data}
\label{rovib_data}
The Spitzer apertures at positions 2 and 11 in the NGC\,1275 nebula
cover similar sky positions to the $K-$band United Kingdom
Infrared Telescope (UKIRT) CGS4 long-slit observations of the `horseshoe
knot' and the `SW1' region of \citet{Hatch05}. Noting that 
the entrance apertures for the Spitzer spectrographs are a different
shape and size to the region extracted from the CGS4 slit we will later use
the P$\alpha$ and H$_2$ ro-vibrational line fluxes of these two
regions from \citet{Hatch05} together with the Spitzer mid-infrared
H$_2$ rotational line fluxes of positions 2 and 11 to investigate the
molecular hydrogen in these outer regions of the nebula. Full details
of the observations and data reduction of the $K$-band data from these
regions are provided in \citet{Hatch05}.

The Spitzer aperture covering the East region does not match well to
the Eastern slit region of \citet{Hatch05}. Instead we have used
UKIRT UIST $HK-$band longslit spectra of the bright radial filament
that extends from a region 2.65\,arcsec South of the galaxy nucleus
eastward to a distance of 26.8\,arcsec (9.5\,kpc). The observations
were obtained on 2005 January 31, and the sky conditions were
photometric.  The long-slit was used in a 7\,pixel-wide (0.84\,arcsec)
configuration with a spatial scale of 0.1202\,arcsec\,pixel$^{-1}$,
using the $HK$ grating which gives a spectral coverage of
$1.4-2.5\mu$m at a resolution of $700-1260$\kmps. The observations were taken
in NDSTARE mode with an object-sky-sky-object nodding pattern. Details
of the observation are provided in Table \ref{tab:UIST_log}. Sky
emission features were mostly removed by the nodding pattern, while flux
calibration was achieved using a set of almost featureless F and G
type stars of known magnitude. In each of the standards a Br$\gamma$
stellar feature at 2.166$\mu$m was removed by linear interpolation.
The data were reduced with ORAC-DR available through the UKIRT
Website, and Starlink packages.

\begin{table}
\caption{Details of the UIST long-slit observation of
  the NGC~1275 East region. The orientation of the nodding direction
  is described in the UIST manual$^1$.
Position angle is measured East from North.}
\begin{tabular}{lllll}
\hline
  Target position&Slit angle&Nod offset&Exposure &Seeing \\
  (J2000)& (degrees)&(arcsec)&time (mins)&(arcsec)\\ \hline
  03 19 48.16 &+70 &+38, -23&144&0.4\\ 
  41 30 40.1\\ 
  \hline
\end{tabular}
\label{tab:UIST_log} 
\end{table}

\footnotetext[4]{\url{http://www.jach.hawaii.edu/UKIRT/instruments/uist/spectroscopy/spectroscopy.html}}

\begin{figure}
\centering
\includegraphics[height=1\columnwidth, angle=-90]{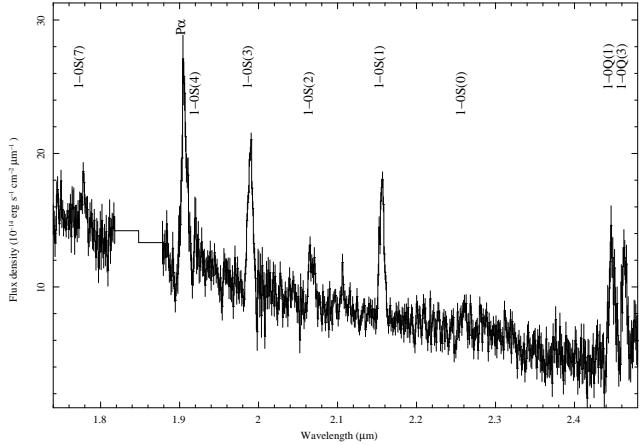}
\caption{Near-infrared spectrum of the NGC~1275 East region. The region between
  1.8$-$1.88$\mu$m has poor atmospheric transmission and has been
  masked out of the spectrum for clarity. Errors are from Poisson statistics.}
\label{fig:HK-east-spectrum}
\end{figure}

A spectrum was extracted from the UIST
long-slit data by summing the flux across the entire region
where the more northerly Spitzer aperture overlaps with the UIST
aperture. This results in a total long-slit aperture of
0.84$\times$6.6\,arcsec$^2$.
Fig.~\ref{fig:HK-east-spectrum} shows the NGC~1275 East
near-infrared spectrum over the wavelength range $1.74-2.48\mu$m. The
wavelength range 1.8$-$1.88$\mu$m is a region of poor atmospheric
transmission and has been masked out of the spectrum for clarity of
the emission-line features. The molecular hydrogen emission lines and
P$\alpha$ are clearly visible. Table \ref{tab:Per_east_fluxes_nir} lists
the surface brightnesses of all the emission-lines.
These are lower limits to the intrinsic surface brightnesses
because of the unknown
covering factor of the excited molecular hydrogen gas.

\begin{table}
\centering
\caption{Near-infrared line surface brightnesses for the NGC~1275 East
  region together with 1$\sigma$ uncertainties. The value for the $2-1$
S(1) line is a 3$\sigma$ upper limit.}
\begin{tabular}{lllllllllll}
\hline

Line & Surface Brightness \\
     & $\times10^{-16}\ergpscmpspsas$\\
\hline
P$\alpha$       & 1.73$\pm$0.08\\
H$_2 1-0$\,S(0) & 0.24$\pm$0.05\\
H$_2 1-0$\,S(1) & 1.26$^{+0.08}_{-0.03}$\\
H$_2 1-0$\,S(2) & 0.55$\pm$0.05\\
H$_2 1-0$\,S(3) & 1.43$\pm0.05$\\
H$_2 1-0$\,S(4) & 0.23$\pm$0.08\\
H$_2 1-0$\,S(7) & 0.35$^{+0.03}_{-0.04}$\\
H$_2 1-0$\,Q(1) & 1.17$\pm$0.11\\
H$_2 1-0$\,Q(3) & 1.09$\pm$0.11\\
H$_2 2-1$\,S(1) & $<$0.12\\

\hline
\end{tabular}
\label{tab:Per_east_fluxes_nir}
\end{table}

\section{Analysis}
\label{analysis}

The lines detected in the Spitzer mid-infrared spectra consist of
emission from molecular hydrogen, from various atomic species of
Neon, Sulphur and Silicon and from the 11.3 micron PAH feature.

\subsection{Diagnostic line ratios}
\label{diagnosticratios}
\begin{figure}
\protect\resizebox{\columnwidth}{!}
{
\includegraphics[angle=0]{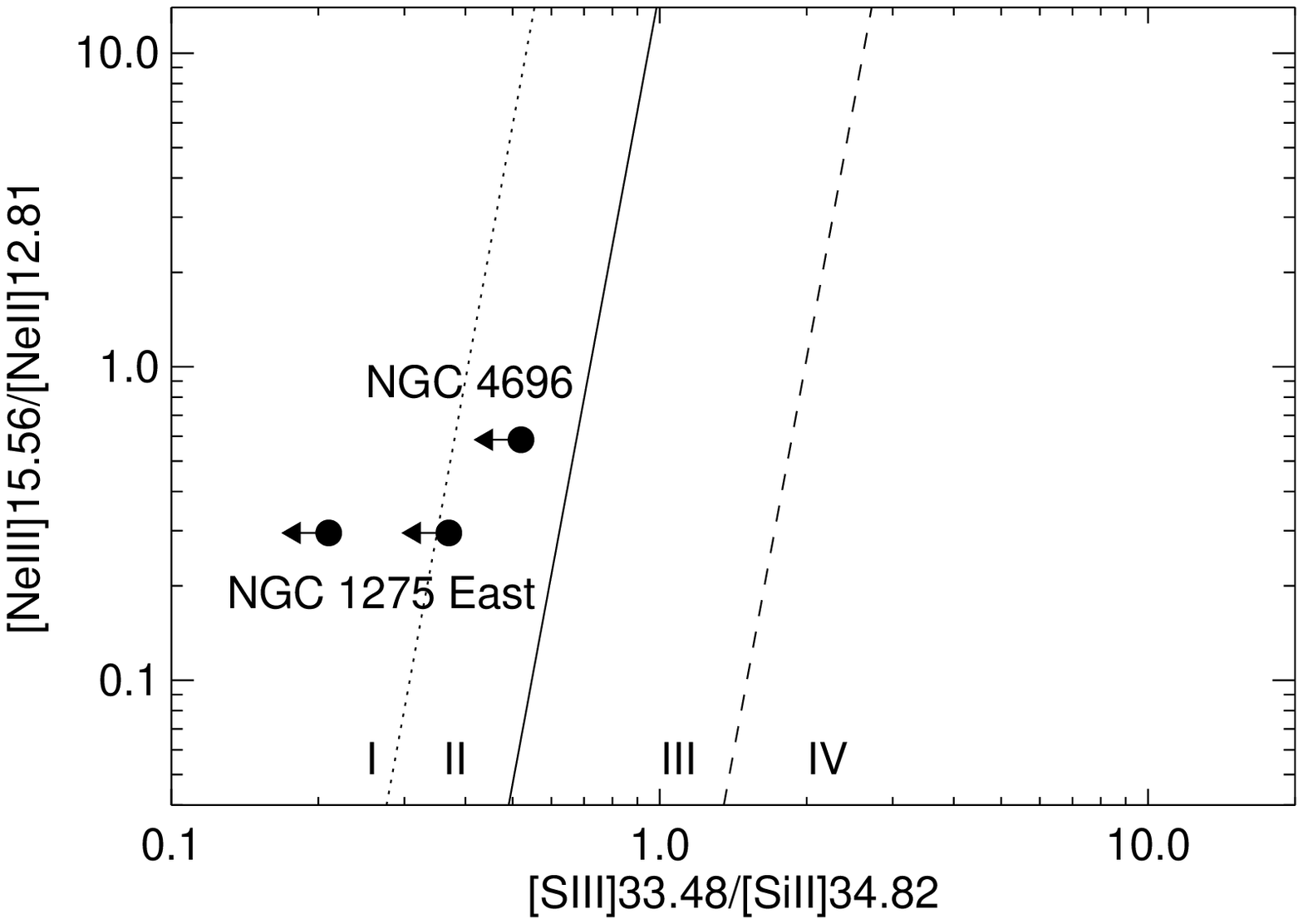}
}
\caption{Neon/Sulphur/Silicon diagnostic diagram of
  \citet{Dale06}. Regions I+II demark the locus of Seyfert and Liner
  nuclei. Region III demarks the locus of HII nuclei while Region IV
  is the locus of Extranuclear and  HII regions.}
\label{fig:daleplot}
\end{figure}

In order to classify the emission-line regions seen by Spitzer, we have
plotted the lines of Neon/Sulphur/Silicon in  Fig~\ref{fig:daleplot},
on one of the diagnostic diagrams of \citet{Dale06}.
The NGC~1275 East and NGC~4696 points both lie in the Region I+II area of the
diagram occupied by Seyfert and Liner nuclei. This
finding is consistent with other analyses of optical emission lines in these
objects which also show Liner-like spectra (eg \citealt{Johnstone87,Crawford99}).

\subsection{[NeIII] lines}
\label{neiiianal}
The SH spectra of the NGC~1275 East region and the NGC~4696 pointing
both show strong detections of the [NeIII] $\lambda15.56\mu$m line. In
the NGC~1275 East region this is a surprising discovery since most of
the very extended emission-line gas does not show emission from the
optical lines of [OIII]$\lambda\lambda$4959,5007 \citep{Hatch06}.  The
ionization potential of Ne$^{+}$ (41.07 eV) is larger than the
ionization potential of O$^{+}$ (35.11eV) so if lines of [NeIII] are
seen we would expect to also see lines of [OIII].

There are two possible explanations for the apparent lack of the
[OIII] lines when the [NeIII] lines are strong, assuming that they are
both formed in a photoionized nebula. The first possibility is that
the [OIII] and [NeIII] lines are both emitted together from a spatial
region which is missed in our optical long-slit spectra but sampled in
the Spitzer spectra due to the much larger SH aperture
($4.7\times11.3$ arcsec) and the poorer point-spread function of
Spitzer.  Alternatively, since these lines are collisionally excited
and the excitation energy of the upper level in the [OIII] lines
($\sim$29,000K) is much greater than that for the upper level in the
[NeIII] line ($\sim$940K), if the electron temperature in the emitting
gas is low enough, the infra-red [NeIII] line may be produced while
the optical [OIII] lines are not excited.

To determine whether the first possibility is occurring we note that
the SH aperture for our NGC~1275 East observations is located very
close to an HII region identified by \citet{Shields90}. That HII
region has much stronger [OIII] emission than the surrounding nebula
and therefore may contribute significant [NeIII] emission.

In order to assess whether the Spitzer SH apertures contain a
significant amount of light from the HII region we used the Spitzer
Stinytim point-spread function modelling software to generate an image
of the point-spread function for the SH instrument at the observed
wavelength of [NeIII] ($15.8\mu$m). We then applied a world coordinate
system to this image such that the centre of the point spread
function lies at the coordinates of the HII region and the pixel scale
is correct for the point-spread function image. The Spitzer aperture region
files corresponding to the two exposure ids were then applied to this
image and the counts in each region were measured.

This analysis predicts that the region corresponding to exposure id
0002 (the more westerly one, nearer the HII region) should have 2.86
times the flux present in the region corresponding to exposure id 0003
if the flux comes entirely from the point source HII region.  The
ratio of fluxes measured in the [NeIII] lines in these two exposure
ids is just 1.23. If we further assume that each exposure id has a
contribution from the extended (non-HII region) nebula proportional to
the H$\alpha$+[NII] emission listed in Table \ref{tab:hafluxes}, then
we can solve for the fractional contribution of the HII region to the
[NeIII] flux in each exposure id. Under these assumptions, we find
that 3 per cent of the [NeIII] flux ($1.5\times10^{-16}\ergpscmps$) in
exposure id 0002 comes from the HII region while only 1 per cent of
the [NeIII] flux ($5.4\times10^{-17}\ergpscmps$) in exposure id 0003
comes from the HII region. The [NeIII] flux from the extended nebula
is $5.25\times10^{-15}\ergpscmps$ and $4.35\times10^{-15}\ergpscmps$ in
exposure ids 0002 and 0003 respectively.  This suggests that there is
a strong extended [NeIII] emission component in the NGC~1275 East region.

We have set a $3\sigma$ upper limit on the flux in the
[OIII]$\lambda5007$ line by extracting a spectrum from the Gemini data,
presented by \citet{Hatch06}, close to the Perseus East region (and
avoiding the HII region).  In order to avoid having to correct for
different spatial region sizes between the optical and infra-red data
we proceed by taking ratios of the [OIII] and [NeIII] lines to the
H$\alpha$+[NII] lines:
$$\frac{\rm{F}(5007)}{\rm{F}(15.56)}=\frac{\rm{F}(5007)}{\rm{F}(\rm{H}\alpha+[\rm{N}II])}
/\frac{\rm{F}(15.56)}{\rm{F}(\rm{H}\alpha+[\rm{N}II])}$$
\noindent where the ratio
$\rm{F}(5007)/\rm{F(\rm{H}\alpha+[\rm{N}II])}$ is the ratio of
the flux in the [OIII]$\lambda$5007 line to the flux in the sum of the
H$\alpha$+[NII]$\lambda\lambda6548,6584$ lines both measured from the
Gemini spectrum. The ratio
$\rm{F}(15.56) / \rm{F}(\rm{H}\alpha+[\rm{N}II])$ is measured
from the [NeIII] flux in the Spitzer spectrum and the values given in Table
\ref{tab:hafluxes}. We find a $3\sigma$ upper limit on the ratio
$\rm{F}(5007)/\rm{F}(15.56) < 1.2$.

Gas that is photoionized by a stellar or power law continuum generally
has a temperature around 10,000 K (\citealt{Osterbrock06}).  The lack
of optical [OIII] emission at a position where infrared [NeIII]
lines are seen suggests that the electron temperature is low enough to
prevent optical emission from being produced. Models using
the code Cloudy (\citealt{Ferland98}) show that if the abundance of
O$^{++}$/Ne$^{++}$ has the solar ratio, then the ratio
$\rm{F}(5007)/\rm{F}(15.56) \sim 10^4T_e^{-1/2}\rm{e}^{(-29000/T_e)}$.
The observed limit of 1.2 in this ratio therefore corresponds to $T_e<6300$K.

Low temperatures in a photoionized gas are produced when the
abundances are at or above solar and the gas density is low enough
that most of the cooling luminosity is radiated by the fine structure
lines (\citealt{Ferland84, Shields95}). These calculations show that
metallicities in the range of solar to twice solar can produce a drop
in the [OIII] line strength by up to 2 orders of magnitude depending
on the details of the physical system.

The results from the discussion above rely on the observed optical
line ratios (which have been corrected for reddening within our
Galaxy) not being affected by any obscuration within the emission line
regions themselves. The Balmer decrement, calculated from the ratio of
H$\alpha/$H$\beta$ in the Gemini spectrum, is very high (6.7) even
after correcting for Galactic reddening. If this high value of the
Balmer decrement is attributed to extinction in a screen in front of
the emission-line region, then the implied extinction is
A$_V$=2.67. After correcting the optical lines for this amount of
reddening we obtain a limit on the ratio
$\rm{F}(5007)/\rm{F}(15.56) < 19$ which would not set an
interesting limit on the electron temperature in the nebula. We note
that high values of the Balmer decrement may be produced by mechanisms
other than reddening, for example collisional excitation (Table 5,
\citealt{Parker64}).

In NGC~4696, \citet{Johnstone88} report [OIII]$\lambda$5007/H$\beta
\sim 0.2$ in a $13.2\times 1.5$ arcsec region at position angle
$85^\circ$ centred 5 arcsec south of, but including, the nucleus,
while \citet{Lewis03} set an upper limit of
[OIII]$\lambda$5007/H$\beta < 0.4$ in a region $1.8\times1.0$ arcsec
centred on the nucleus. This indicates that the [OIII] line is quite
weak in this object also. We do not attempt to calculate the
$\rm{F}(5007)/\rm{F}(15.56)$ ratio due to the much larger
uncertainties because of the proximity of the Spitzer aperture to the
galactic nucleus (Fig. \ref{fig:cenmap}).  The inner region of the
Centaurus cluster is known to have a high metallicity from X-ray
observations (eg \citealt{Sanders06}), although the exact value
derived for the very central regions is dependent on the precise model
fitted. If future observations were to be able to constrain the
metallicity of the emission-line gas, that information might prove an
interesting constraint on the origin of the emission-line gas.

\subsection{Density diagnostic}
\label{densitydiagnostic}
We note that the [SIII]$\lambda$18.71/[SIII]$\lambda$33.48 line ratio
is a density diagnostic. In the NGC~1275 East and NGC~4696 pointings
we have detections of $\lambda18.71$ and upper limits for
$\lambda$33.48. Taking the ratio of the average of the
line fluxes (or upper limits), having used the H$\alpha$+[NII] fluxes in the Spitzer apertures
to scale the fluxes to account for the different spatial sampling as in section
\ref{spitzer_data}, gives [SIII]$\lambda$18.71/[SIII]$\lambda$33.48 $>0.6$ for NGC~1275 East. Cloudy
(\citealt{Ferland98}) models show that this ratio corresponds to a
density of $>100\pccm$. No constraint is
placed on the density for the NGC~4696 pointing as we are unable to
scale the fluxes for the aperture effects due to saturation in the
H$\alpha$+[NII] image.

\subsection{PAH features}

11.3~$\mu$m PAH features are detected in the spectra of NGC~1275 East
and NGC~4696. The flux is strong and comparable to that of [NeII].
No detection is made for Perseus positions 2 and 11 which are at
greater distances from the centre of the galaxy than Perseus East. The
limit on position 2 is less than about one third that of [NeII]
indicating that either the abundance of PAHs or the excitation
mechanism is much reduced there.  \citet{Peeters04} argue that PAH
features are tracers of star formation, particularly of B stars. Parts
of NGC\,1275 have an A-type spectrum \citep{Minkowski68} which
indicates some, possibly sporadic, star formation; the situation in
NGC\,4696 is uncertain due to the obvious dust lanes. The correlation
between the presence of a PAH feature and a stellar continuum in the
IRS spectrum in NGC\,1275 supports an excitation mechanism involving
the far UV light from moderately young stars.

In a Spitzer study using IRS and Multiband Imaging Photometer (MIPS)
observations of 7 nearby dusty elliptical galaxies including
NGC~4696, \citet{Kaneda07} find PAH emission from the centre and
evidence for by dust emission which is more extensive than the
starlight of the galaxy in the MIPS data.

\subsection{Molecular Hydrogen Lines}
To investigate the molecular hydrogen in more detail we can combine the
mid-infrared pure rotational emission lines with emission lines from the
1-0\,S(J) ro-vibrational states visible in the $K-$band (Section \ref{rovib_data})

\subsubsection{Excitation process}
Molecular hydrogen can be collisionally excited, or radiatively
excited by absorbtion of UV photons in the Lyman-Werner bands.  If the
density is low enough ($<10^{4}$cm$^{-3}$) the molecule will de-excite
through the ground electronic ro-vibrational states by radiative
de-excitations, whereas if the density is high, the populations of the
rotational and vibrational states will be redistributed by collisions
before de-excitation to the ground level. The form of the observed
spectrum depends on the excitation source as well as the density and
temperature of the gas. At high temperatures and densities the
collision rate of the H$_2$ molecule with other molecules and atomic
species increases. Collisional de-excitation therefore becomes more
important than radiative de-excitations as the temperature and density
increase. The total gas density at which the collisional de-excitation
rate equals the radiative de-excitation rate is
known as the critical density and is a slow function of
temperature \citep{Sternberg89,Mandy93}. When collisional
excitation and de-excitation dominates, the gas exists in local
thermodynamic equilibrium (LTE) and the molecular hydrogen has an
ortho-to-para ratio of three.

\begin{table}
\centering
\caption{The intensity ratio of H$_2$ $2-1$\,S(1)/$1-0$\,S(1) lines in the three regions of NGC\,1275. The $2-1$\,S(1) line fluxes are 3$\sigma$ upper limits
  as the line is not detected in any region.}
\begin{tabular}{lc}
\hline
Region&$2-1$\,S(1)/$1-0$\,S(1)\\ \hline
  NGC~1275 East& $<$0.097\\
  Position 2 &$<$0.26\\
  Position 11&$<$0.44\\ \hline
\end{tabular}
\label{2-1limits}
\end{table}

\subsubsection{Scaling the mid-infrared H$_2$ emission to compare with
  the near-infrared emission}
\label{sec:scaling}
\begin{figure}
\protect\resizebox{\columnwidth}{!}
{
\includegraphics{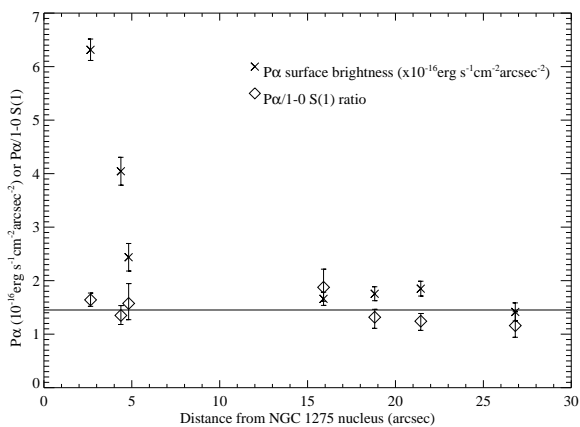}
}
\caption{P$\alpha$ flux and P$\alpha$/$1-0$\,S(1) ratio along the UIST
  long-slit covering the NGC~1275 East region. The solid line shows the mean
  P$\alpha$/$1-0$\,S(1) of 1.45. Error bars are at 1$\sigma$ level.
  The ratio of atomic-to-molecular hydrogen line emission is fairly
  constant across a large range of P$\alpha$ emission.}
\label{fig:pa-s1}
\end{figure}

As discussed in Section \ref{spitzer_data} we have assumed that the
mid-infrared line emission is distributed spatially like the H$\alpha$+[NII]
emission. The near-infrared H$_2$ line emission has been shown to
closely follow the P$\alpha$ emission in other brightest cluster
galaxies \citep{Jaffe05}. Additionally, near-infrared H$_2$ emission
lines are only detected in NGC\,1275 in the regions of brightest
H$\alpha$ emission \citep{Hatch05}. To examine the relationship
between the atomic hydrogen emission and the near-infrared molecular
hydrogen emission within NGC\,1275 in detail, we have analysed the
whole length of the NGC~1275 East $HK$-band long-slit data. The long-slit
data were segregated into seven regions in which P$\alpha$ was dectected.
Fig.~\ref{fig:pa-s1} plots the P$\alpha$ flux and P$\alpha$/$1-0$\,S(1)
line ratio against projected distance from the galaxy nucleus for all
seven regions. This figure shows that the P$\alpha$/$1-0$\,S(1) ratio is
almost constant at 1.45 across a large part of the emission-line region. We
therefore assume that the P$\alpha$ emission and near-infrared H$_2$
line emission are closely related spatially in NGC\,1275. We
can scale and combine the near-infrared and mid-infrared H$_2$
emission lines using the flux of the hydrogen recombination lines in
the Spitzer and near infra-red apertures.

To obtain Hydrogen line fluxes for the different spectra in different
regions we proceed as follows: For the Spitzer spectra we used the
H$\alpha$+[NII] fluxes from the figures given in Table \ref{tab:hafluxes}
divided by two to account for the [NII] lines \citep{Hatch06}. For
the near infra-red spectra we used the P$\alpha$ flux from Table
\ref{tab:Per_east_fluxes_nir} for the Perseus East region, the
P$\alpha$ flux given by \citet{Hatch05} for the Position 2 region and
$1.45\times$ the flux in the H$_2\,1-0$\,S(1) line (Fig \ref{fig:pa-s1}) for
the Position 11 region. This corresponds to the `Horseshoe knot'
in \citet{Hatch05} which does not have a measured P$\alpha$ intensity
as the line was badly affected by the poor atmospheric transmission in
that part of the spectrum. To scale the P$\alpha$ flux to the
H$\alpha$ flux we assume Case B recombination theory
\citep{Osterbrock06} which implies that H$\alpha$/P$\alpha = 8.45$. We
finally scale the Spitzer line fluxes to match the near infra-red
spatial regions with:

$$\rm{F}_{\rm ss}={\rm F}_{\rm sm}\times \frac{{\rm F}_{\rm nir}({\rm
    H}\alpha)} {{\rm F}_{\rm s}({\rm H} \alpha)}$$

\noindent
where: F$_{\rm ss}$ is the resulting scaled Spitzer line flux, F$_{ \rm sm}$ is
the measured Spitzer line flux, F$_{\rm nir}$(H$\alpha$) is the estimated
H$\alpha$ flux in the near infra-red aperture, F$_{\rm s}$(H$\alpha$)
is the estimated H$\alpha$ flux in the Spitzer aperture.

\subsubsection{Excitation temperatures, column densities and mass of
  warm molecular hydrogen}
If collisions share the energy between the
particles causing the H$_{2}$ to be in LTE,
the excited states of H$_{2}$ ($N(\vib{v}, J)$) will be
populated in a thermalised Boltzmann distribution characterised by an
excitation temperature (T$_{{\rm ex}}$) such that:
\begin{equation}
 \frac{N(\vib{v},J)}{g_{J}}= a(T_{ex})~e^{-E(\vib{v},J)/{\rm k_{B}}T_{{\rm ex}}}~,
\label{equation:t_ex}
\end{equation}
where $E(\vib{v}, J)$ is the upper energy of the $(\vib{v},J)$
transition, $a(T_{ex})$ is equal to $N_{\rm Total}/Z_T$, where
$N_{\rm Total}$ is the total H$_2$ column density, $Z_{T} $ is the
H$_2$ partition function, and g$_{J}$ is the statistical weight.
$N(\vib{v}, J)$ is the column density of the $(\vib{v},J)$ level,
which, for optically-thin emission can be observationally determined
from $I$ (the surface brightness
[erg\,s$^{-1}$cm$^{-2}$sr$^{-1}$]) through:
\begin{equation}
  N(\vib{v}, J)=\frac{4 \pi \lambda I}{{\rm A_{ul}} h c}~,
\label{equation:intensity}
\end{equation}
where $\lambda$ is the rest wavelength of the line and A$_{\rm{ul}}$ is the
Einstein co-efficient taken from \citet{Turner77}. If the H$_{2}$ level
populations are completely dominated by collisional excitation
and de-excitation then the excitation temperature equals the kinetic
temperature of the gas.

Population diagrams of
$\ln[\frac{N(\vib{v}, J)}{g_{J}}]$ verses $E(\vib{v}, J)/{\rm
  k_{b}}T_{{\rm ex}}$ allow investigation of whether the levels are
thermalised. If the levels are thermalised the level populations will
lie in a straight line with a slope inversely proportional to the
excitation temperature. If a range of gas temperatures occur along the
line-of-sight, the points will lie on a smooth curve with the lower
energy levels lying on a steeper slope (lower temperature) than the
higher energy levels (higher temperature).
This is because the higher energy levels are preferentially
populated at higher temperatures. 

\begin{figure}
\protect\resizebox{\columnwidth}{!}
{
\includegraphics{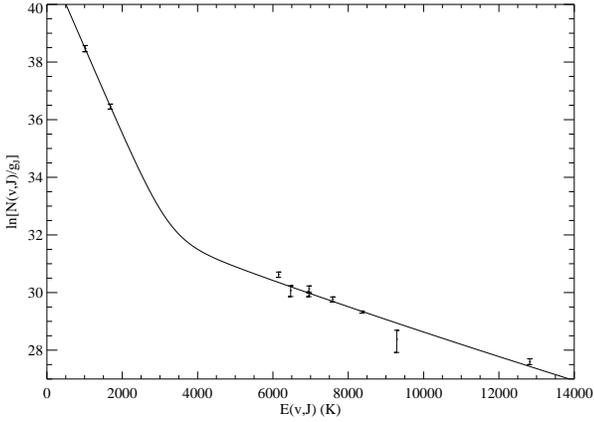}
}
\caption{Population diagram of the NGC~1275 East region with 1$\sigma$
  error bars. The datapoints at $E=1015$\,K and 1682\,K are
  measured by Spitzer and the error bars include the uncertainty in
  scaling the mid-infrared emission to the near-infrared emission
  using hydrogen recombination lines.}
\label{pop_east}
\end{figure}

\begin{figure}
\protect\resizebox{\columnwidth}{!}
{
\includegraphics{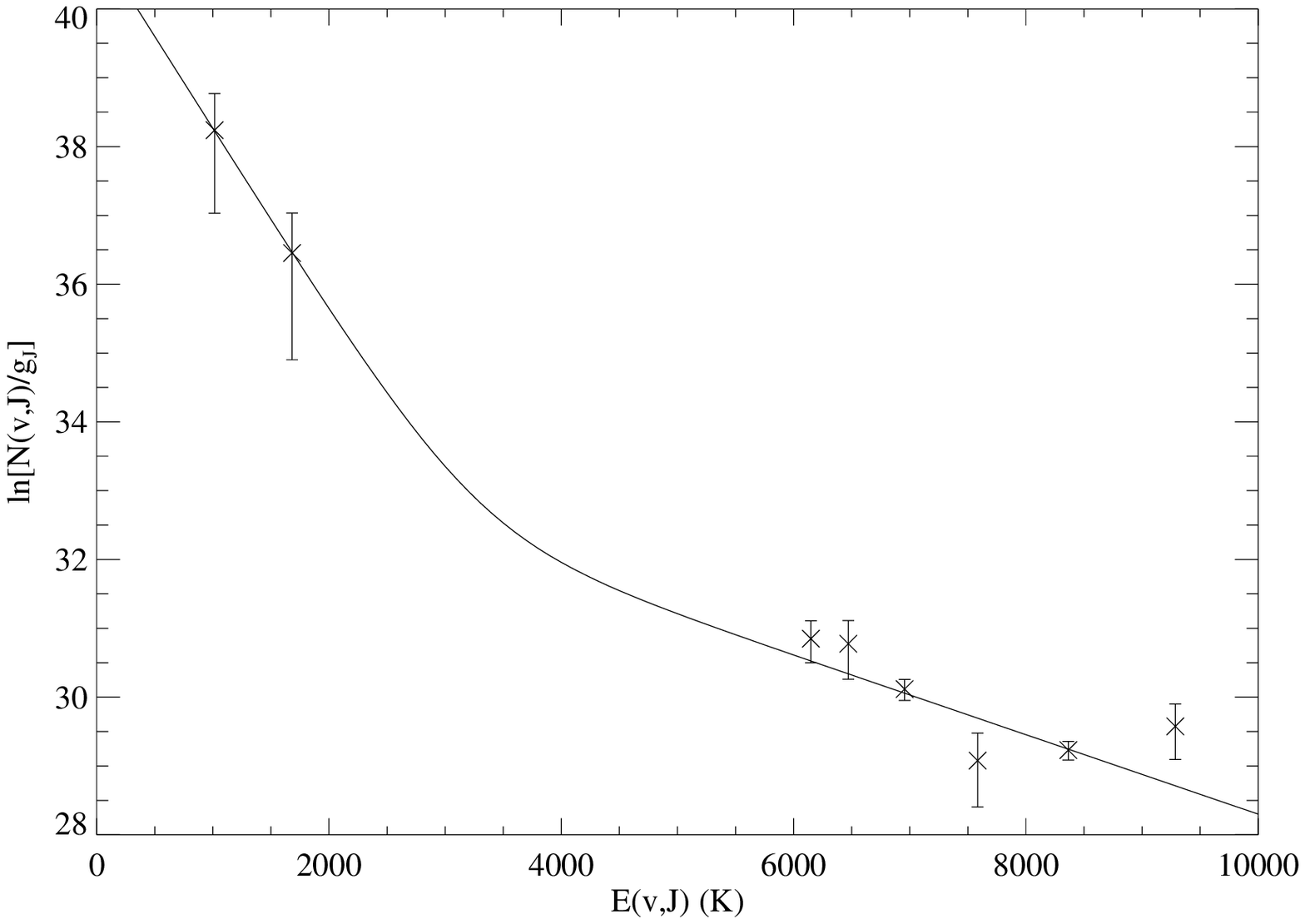}
}
\caption{Population diagram of NGC~1275 position 2 with 1$\sigma$ error
  bars. The datapoints at $E=1015$\,K and 1682\,K are
  measured by Spitzer and the error bars include the uncertainty in
  scaling the mid-infrared emission to the near-infrared emission
  using hydrogen recombination lines.}
\label{pop_2}

\end{figure}

\begin{figure}
\protect\resizebox{\columnwidth}{!}
{
\includegraphics{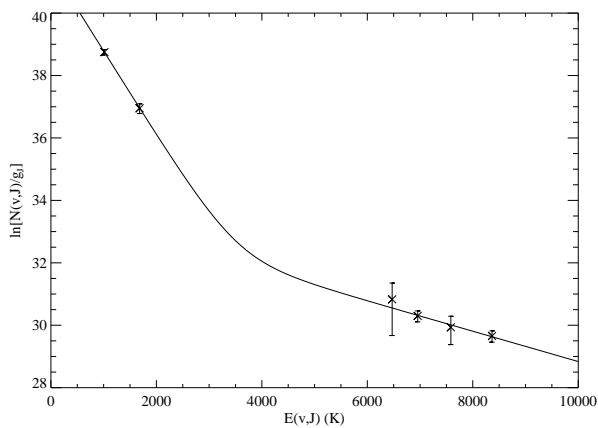}
}
\caption{Population diagram of NGC~1275 position 11 with 1$\sigma$
  error bars. The datapoints at $E=1015$\,K and 1682\,K are
  measured by Spitzer, but the error bars do not include any
  uncertainty associated with scaling the mid-infrared emission lines.}
\label{pop_11}
\end{figure}

Population diagrams for the NGC~1275 East, position 2 and position 11
regions are presented in Figs.~\ref{pop_east}, \ref{pop_2} and
\ref{pop_11}. As the population of the states lie on a smooth-curve,
these diagrams clearly show that the molecular hydrogen exists in LTE
at a range of temperatures. Supporting evidence that the H$_2$ is
thermally excited comes from the ratio of the H$_2$
$2-1$\,S(1)/$1-0$\,S(1) line intensity ratio. 3$\sigma$ upper limits
for the $2-1$\,S(1) emission lines were measured from the position 2,
11 and the NGC~1275 East regions and the limits on the
$2-1$\,S(1)/$1-0$\,S(1) ratios are given in Table \ref{2-1limits}. The
measured ratio is significantly less than the pure radiative
de-excitation value of 0.53 \citep{Mouri94}, so the gas is
predominately de-excited through collisions, perhaps with a small
non-thermal component.  This implies that the gas must have a density
of greater than $10^{4}$cm$^{-3}$.

To gain a simple picture of the structure of the molecular gas we have
measured the rotational temperature from the $0-0$\,S(1) and
$0-0$\,S(2) lines, and a ro-vibrational temperature from the $1-0$
lines. In the NGC~1275 East $HK$-band spectrum we measured the
$1-0$\,S(7) line, and therefore we measure two ro-vibrational excitation
temperatures in this region: one from levels with an upper energy of
$<8000$\,K and one from levels with an upper energy of $>8000$\,K.
These excitation temperatures are given in Table \ref{temp_den}. The
pure-rotational temperatures of the gas are $330-370$\,K for all
regions.  The ro-vibrational lines reveal a layer of hotter gas at
$1700-2000$\,K and in NGC~1275 East there is an additional layer of
even hotter gas at $\sim2600$\,K. Such high population temperatures
are often interpreted as due to non-thermal excitation, for instance
by starlight \citep{vanDishoeck04} or cosmic rays \citep{Dalgarno99}.
In the remainder of this paper we will refer to the high v,J
populations as an indication of warm H$_2$, without meaning to specify
what actually causes these level populations.

\begin{table*}
\caption {Excitation tempertures, column densities and masses of
  molecular Hydrogen. The observed mass is calculated directly from
  the observed flux in the relevant aperture. The scaled mass is the mass
scaled to the region in which
the ro-vibrational lines were observed. Error bars are at the
1$\sigma$ level.}
\begin{tabular}{llll}
\hline
  & rotational &ro-vibrational &ro-vibrational\\ 
  &  emission lines&($2000<$E$_{\rm U}<8000K$)&(E$_{\rm U}>8000K$)\\ \hline
  NGC~1275 East & & &\\
T$_{\rm ex}$&330$\pm$20\,K&1730$\pm^{280}_{210}$\,K&2580$\pm^{160}_{145}$\,K\\
Column density&4.2$\times$10$^{18}$cm$^{-2}$&4.4$\times$10$^{15}$cm$^{-2}$&3.4$\times$10$^{15}$cm$^{-2}$\\
Observed mass of H$_2$&$1.5\times10^6\Msun$&135$\Msun$&55$\Msun$\\
Scaled mass of H$_2$&$5.0\times10^4\Msun$&135$\Msun$&55$\Msun$\\
\hline

  NGC~1275 Position 2 & & &\\
T$_{\rm ex}$&370$\pm$20\,K&1730$\pm^{310}_{230}$\,K&\\
Column density&2.6$\times$10$^{18}$cm$^{-2}$&1.3$\times$10$^{16}$cm$^{-2}$&\\
Observed mass of H$_2$&$2.9\times10^5\Msun$&131$\Msun$&\\
Scaled mass of H$_2$&$2.6\times10^4\Msun$&131$\Msun$&\\
\hline

  NGC~1275 Position 11 & & &  \\
T$_{\rm ex}$&370$\pm^{40}_{30}$\,K&2060$\pm^{950}_{490}$\,K&\\
Column density&4.5$\times$10$^{18}$cm$^{-2}$&1.1$\times$10$^{16}$cm$^{-2}$&\\
Observed mass of H$_2$&$1.5\times10^5\Msun$&53$\Msun$&\\
Scaled mass of H$_2$&$2.2\times10^4\Msun$&53$\Msun$&\\
\hline

  NGC~4696 & & &  \\
T$_{\rm ex}$&310$\pm10$\,K&\\
Column density&3.5$\times$10$^{18}$cm$^{-2}$\\
Observed mass of H$_2$&$1.3\times10^5\Msun$\\
\hline

\end{tabular}
\label{temp_den}
\end{table*}

The total H$_{2}$ column density of the molecular hydrogen can be
calculated through rearranging equation \ref{equation:t_ex} and
assuming that all the molecular hydrogen exists in LTE
at the excitation temperature derived from the population
diagrams.
\begin{equation}
N_{{\rm Total}}=\frac{N(\vib{v}, J)Z(T)}{g_{J}e^{-E(\vib{v}, J)/{\rm k_{B}}T_{{\rm ex}}}}~,
\label{equation:columndensity}
\end{equation}
where $Z(T)$ is the partition function.

As the molecular hydrogen exists at multiple temperatures we have
assumed the gas consists of two (or three in the case of NGC~1275 East)
separate populations, each consisting of a Boltzmann distribution at
the excitation temperatures derived from the population
diagrams. A least squares fit was performed with the total column
densities of each Boltzmann distribution as the two (or three) free
parameters. Column densities at each excitation temperature for the
three regions are given in Table \ref{temp_den}.

The mass of warm molecular hydrogen within the Spitzer aperture can be
derived from a single line luminosity and the excitation temperature.
The total molecular hydrogen mass ${\rm M}_{\rm Total}$ is derived
from:

\begin{equation}
{\rm M}_{\rm Total}={\rm M}_{{\rm H}_2}{\rm n_{\rm total}},
\end{equation}
where ${\rm M}_{{\rm H}_2}$ is the mass of a hydrogen molecule, and
${\rm n_{\rm total}}$ is the total number of H$_2$ molecules.  ${\rm
  n_{\rm total}}$ can be derived from the luminosity from the ($\vib{v},J$) state
through:

\begin{equation}
{\rm n_{\rm total}}=\frac{L(\vib{v}, J)Z_{T_{ex}}}{{\rm A}_{\rm ul}h\nu g_{J}e^{-E(\vib{v}, J)/{\rm K_B}T_ex}},
\end{equation}
where $L(\vib{v}, J)$ is the luminosity in the $(\vib{v}, J)$ line,
$Z_{T_{ex}}$ is the partition function, ${\rm A_{\rm ul}}$ is the
Einstein co-efficient, and $\nu$ is the frequency of the line. The
masses of the $\sim$350\,K material in each of the Spitzer apertures
were calculated from the 0-0\,S(1) line. The
masses of the warmer gas (1700--2000\,K) were measured using the
1-0S(1) line. A check was made using the other detected lines
1-0S\,(2), 1-0S\,(3), 1-0S\,(4) and the masses derived from these
lines were within 10 per cent of the massess derived from the 1-0S(1)
line. Finally the mass of the 2580K material seen in the NGC\,1275
East region was measured from the 1-0\,S(7) line.  The masses of
molecular hydrogen derived from the different temperature lines are
given in Table \ref{temp_den} as the ``Observed mass of H$_2$''. In
the rows labelled "Scaled mass of H$_2$" we list the masses of
molecular hydrogen at each temperature scaled to the region in which
the ro-virational lines were measured; the Spitzer masses were scaled
down by the ratio of H$\alpha$ fluxes in the relative apertures, as
described in Section \ref{sec:scaling}. The majority of the molecular
gas exits at lower temperatures with less than 0.5 per cent of the
total mass being at the higher temperatures.

Without the measurement of the 0-0 S(0) line, it is not possible to
derive the temperature and mass of of the coldest component of the
warm H2 gas. Therefore we use the 3$\sigma$ upper limits on the
$0-0$\,S(0) flux to obtain lower limits on the temperature and upper
limits on the mass of the coldest component of the warm molecular
hydrogen. These limits are displayed for the NGC1275 pointings in
Table \ref{mass_limits}. We cannot carry out the scaling for the NGC~4696 pointing as
the region of the H$\alpha$+[NII] map covered by the LH aperture is
affected by saturation. The minimum allowed
temperature of the coolest component is 150--180\,K, and approximately
10 times more mass may exist at this cool temperature than in
the warmer molecular hydrogen.

\begin{table*}
\centering
\caption{Limits on the temperature and molecular hydrogen masses from the 0-0S(0) lines.
The observed mass limits are calculated directly from the upper limit
to the observed flux in the relevant aperture. The scaled mass is the
mass scaled to the region in which the ro-vibrational lines were
observed.}

\begin{tabular}{llll}
\hline
Region                & Temperature & Observed mass of H$_2$ &Scaled mass of H$_2$\\
\hline
NGC 1275 East         &$>180$ K       &$<2.9\times10^7$\Msun     &$<3.2\times10^5$\Msun \\
NGC 1275 Position 2   &$>155$ K       &$<1.0\times10^7$\Msun     &$<3.8\times10^5$\Msun \\
NGC 1275 Position 11  &$>150$ K       &$<5.6\times10^6$\Msun     &$<3.2\times10^5$\Msun \\
\hline
\end{tabular}
\label{mass_limits}
\end{table*}

\subsubsection{Correlation with Optical Emission Lines}
\begin{figure}
\centering
\includegraphics[height=1\columnwidth, angle=-90]{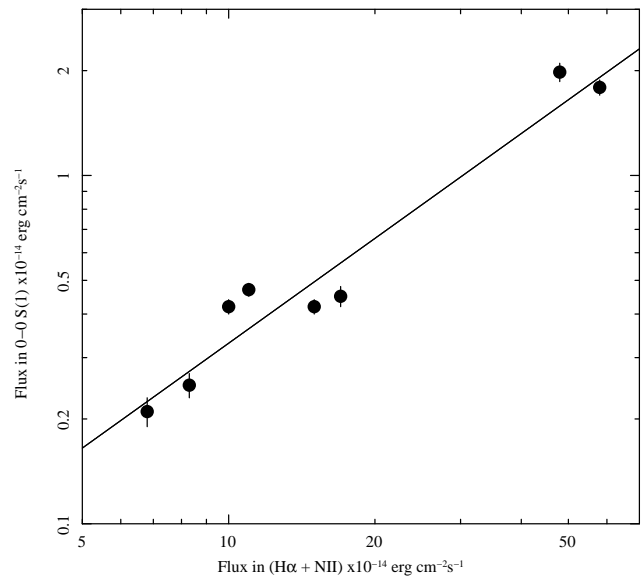}
\caption{Flux in 0-0\,S(1) line plotted against the flux in the
H$\alpha$+[NII] complex for all the NGC~1275 and NGC~4696 pointings.
The solid line is a straight-line fit in linear coordinate space,
constrained to pass though the origin. It has a slope of 0.03.}
\label{fig:h2ha}
\end{figure}

In Fig. \ref{fig:h2ha} we plot the molecular hydrogen 0-0\,S(1) flux
against the H$\alpha$+[NII] flux for all the NGC~1275 and
NGC~4696 pointings. The flux in the
Spitzer 0-0\,S(1) molecular hydrogen line seems to correlate well with the flux
in the optical H$\alpha$+[NII] lines. The fit is a straight
line constrained to go though the origin and shows that the flux in
the 0-0S(1) line is 0.03$\times$ the flux in the H$\alpha$+[NII] complex.

\subsubsection{Pressure balance between the molecular, atomic and X-ray
  emitting gas}
  
There is a serious problem implied by the apparent thermal
distribution of the levels in the molecular hydrogen. This is that the
currently used collisional rates for the ro-vibrational states require
that the density exceeds $10^{4-5}\pccm$ \citep{Sternberg89, Mandy93}.
If the temperature is
few $\times1000$~K then the pressure is $10^{7-8}\pccmK$. This exceeds the
ambient pressure, as determined from the X-ray measurements of the hot
gas (\citealt{Sanders05}), by an order of magnitude or more. This does not
appear to be a stable situation for such thin, apparently long-lived,
filaments.

Recent studies have brought the H$_2$ collision database into
question.  The Orion Bar is the closest and best studied H$^+$ / H$^0$
/ H$_2$ interface (\citealt{Odell01}).  \citet{Allers05} found that
H$_2$ rotational lines implied level populations that were close to a
thermal distribution.  They noted that the density within the Bar is
too low to thermalize the levels involved if the current H$_2$
collision rates are correct.  They suggest that current H - H$_2$
vibrational de-excitation rates should be increased by nearly two
orders of magnitude (their Table 6). If this were the case then H$_2$
level populations could reach a thermal distribution at the low
densities implied by the pressure of the hot gas. Our data are further
evidence that the de-excitation rates may require a large correction.

\subsection{Heating Mechanisms}
A detailed discussion of the heating mechanism of the emission-line
gas in central galaxies in cooling core clusters is beyond the scope
of this paper. In the nuclear regions the situation can be complex
with several different mechanisms at work. These include
photo-ionization from an active nucelus (for NGC~1275) (eg
\citealt{Wilman05}), photoionization by stars (eg
\citealt{Johnstone87,Allen95,Wilman02}) and possibly shocks (eg
\citealt{Jaffe01,Wilman02}).

We expect that the off-nuclear regions which we have targeted in the
observations presented in this paper will be more representative of
the general emission-line nebula and also simpler to understand. We
note that the apparently thermalized nature of the molecular
hydrogen lines argues against their formation in a classical
photodissociation region and that the relation between the
H$\alpha$+[NII] flux and the flux in the molecular hydrogen lines
presented in Fig. \ref{fig:h2ha} could suggest that shocks are
important. However, it is not clear that there would be enough
cloud-cloud collisions in these regions to drive this mechanism.
Shocks also need to be constrained to a fairly narrow
range of velocity, fast enough to produce the optical emission lines, but
not so fast that dust is destroyed. \citet{Jaffe01} have previously
discussed the requirement for the fine-tuning of shock velocities in
trying to explain the ro-vibrational molecular lines in a sample of
cooling core central cluster galaxies.

In a future paper we will explore these difficulties in more detail
and examine the possibility of excitation via cosmic rays which leak
out of the radio lobes and ghost bubbles (eg \citealt{Birzan04, Dunn05}).

\section{Conclusion}
\label{conclusion}

\indent Using high resolution mid-infrared spectra from the Spitzer
Space Telescope we have detected off-nuclear emission lines in
NGC~1275 and NGC~4696. The lines arise from pure rotational
transitions of the Hydrogen molecule, atomic fine-structure
transitions from Ne$^+$, Ne$^{++}$, S$^{++}$ and Si$^+$ and the 11.3
micron PAH feature.

The relative intensities of the molecular hydrogen lines are
consistent with collisional excitation and probe a new region of
temperature space in the filaments at 300-400~K. Pressure equilibrium
with the surrounding intracluster medium requires a substantial
revision to H$_2$ collision rates, as noted by \citet{Allers05}.
The outer filaments around NGC\,1275 therefore have molecular gas at
50~K (CO, from P. Salom\'e et al, in preparation), 300--400~K (H$_2$,
this paper) and 2000--3000~K (H$_2$, \citealt{Hatch05}), atomic gas at
several 1000~K (H$\alpha$, e.g. \citealt{Conselice01}) and $\sim
10^7\K$ (\citealt{Fabian03b}), embedded in the hot intracluster gas at
about $5\times 10^7\K$. Detection of OVI from more central regions
with FUSE (\citealt{Bregman06}) suggests that gas at $\sim 3\times
10^5\K$ is also likely to be present. 

The discovery of the [NeIII] lines is unexpected based on the weakness
of the optical [OIII] lines and suggests that the electron temperature
in the filamentary medium where these lines are formed may be very
low, possibly due to enhanced metallicity. The metal abundance of the
central intracluster medium in cooling cluster cores is enhanced by Type
Ia, and possibly Type II supernovae (see \citealt{Sanders06} and
references therein), so the filaments are enriched if they orginate
from cooling of that gas.

Collectively the mid-IR emission lines radiate about one quarter of
the flux in H$\alpha$, so they play a relatively minor role in the
energy flow in the filaments. However the total molecular hydrogen
emission, including the near IR ro-vibrational lines is comparable to
the emission from H$\alpha$. (H$\alpha$ is about one tenth of the
total line emission: the major emitter is expected to be Ly$\alpha$.)
Nevertheless they reveal more about the low temperature core to the
filaments which is dominated in mass by H$_2$. The total mass in H$_2$
for NGC\,1275, obtained from the CO emission using a standard Galactic
conversion ratio, is $\sim 4\times 10^{10}\Msun$ (Salom\'e et al
2006), most of it near the centre of the galaxy. The mass in H$_2$ at
300~K is about $6\times 10^7\Msun$, assuming that the masses found for
the outer filaments in this paper scale linearly with H$\alpha$
emission, and using the total H$\alpha$ flux of Conselice et al
(2001). The total mass of H$\alpha$ emitting gas is smaller at about
$3\times 10^7\Msun$ and the mass of surrounding soft X-ray emitting
gas, with temperature $\sim 7\times 10^6\K$, is about $10^9\Msun$
(Fabian et al 2006).

Further observations are required to test whether all filaments have
similar composition and properties, but they do appear to be
predominantly molecular.

\section{Acknowledgments}
ACF and CSC acknowledge support by the Royal Society.  RMJ and NAH
acknowledge support by the Particle Physics and Astronomy Research
Council.GJF thanks the NSF (AST 0607028), NASA (NNG05GD81G), STScI
(HST-AR-10653) and the Spitzer Science Center (20343) for support.  We
thank an anonymous referee for comments which improved the paper.

\bsp

\label{lastpage}
\clearpage
\end{document}